\documentclass[aps,preprintnumbers,  amsmath, showpacs]{revtex4}
\usepackage{psfrag}
\usepackage{epsfig}
\usepackage{amsfonts} 
\usepackage{amssymb}  
\usepackage{graphicx} 
\usepackage{amsmath}  
\usepackage{amsmath}  
\usepackage{calligra}
\usepackage{mathrsfs}
\usepackage{tabularx}
\usepackage[latin1]{inputenc}
\usepackage{color}
\usepackage{suffix}
\usepackage{mathtools}
\usepackage{slashed}
\usepackage{ulem}

\DeclareMathAlphabet{\mathcalligra}{T1}{calligra}{m}{n}
\DeclareFontShape{T1}{calligra}{m}{n}{<->s*[2.2]callig15}{}

\newcommand{\bea}{\begin{eqnarray}}
\newcommand{\eea}{\end{eqnarray}}
\newcommand{\beas}{\begin{eqnarray*}}
\newcommand{\eeas}{\end{eqnarray*}}

\DeclarePairedDelimiterX\MeijerM[3]{\lparen}{\rparen}%
{\begin{smallmatrix}#1 \\ #2\end{smallmatrix}\delimsize\vert\,#3}

\newcommand\MeijerG[8][]{%
  G^{\,#2,#3}_{#4,#5}\MeijerM[#1]{#6}{#7}{#8}}

\WithSuffix\newcommand\MeijerG*[7]{%
  G^{\,#1,#2}_{#3,#4}\MeijerM*{#5}{#6}{#7}}

\begin{document}


\title{Speed of Sound for Hadronic and Quark Phases in a Magnetic Field}

\author{E. J. Ferrer and A. Hackebill}
\affiliation{Dept. of Physics and Astronomy, University of Texas Rio Grande Valley, Edinburg 78539, USA, 
Physics Department, CUNY-Graduate Center, New York 10314, USA}


\date{\today}

\begin{abstract}
In this paper we calculate the speed of sound for three phases that may exist inside a magnetized hybrid neutron star at different density regions: A hadronic phase at low densities, quark-matter in the magnetic dual chiral density wave (MDCDW) phase at intermediate densities and a free-quark phase modeled by the MIT bag model at higher densities. It is found that the speed of sound exhibits a non-monotonic behavior, that goes from values smaller than the conformal limit ($c_s^2 < 1/3$) in the hadronic phase, to peak ($c_s^2 > 1/3$) in the MDCDW phase, to finally reach the conformal limit ($c_s^2 \sim 1/3$) at higher densities for quarks in the MIT bag model. Also, the anisotropic speed of sound in the presence of a magnetic field is derived  from first principles. This is a consequence of the anisotropy in the system's pressures produced by the breaking of the rotational symmetry in the presence of a magnetic field. The role played by the lowest Landau level contribution in affecting the speed of sound in the magnetized phases is discussed.
\end{abstract}

\maketitle

\section{Introduction}\label{Intro}

Studies of matter under the extreme conditions of high temperature and/or density are topics of increasing general interest in the physics community due to their significance for heavy-ion-collision experiments and for the astrophysics of compact objects. In both contexts, moreover, the presence of strong magnetic fields is undeniable. In this regard, we have that
according to several numerical simulations, off-central Au-Au collisions at RHIC can lead to field strengths of $10^{18}$ - $10^{19}$ G, while the field can be as large as $10^{20}$ G for the off-central Pb-Pb collisions at the LHC \cite{B_HIC-1}.   On the other hand, neutron stars (NS) also exhibit significant strong magnetic fields. The surface magnetic fields of some radio pulsars are of order $10^{8}-10^{12}$ G \cite{NS-B-1}. There are even some special compact objects called magnetars \cite{NS-B-2} whose surface magnetic fields are of order $10^{14}-10^{15}$ G, which have been inferred from spectroscopic and spin-down studies of soft-gamma ray repeaters (SGRs) and anomalous x-ray pulsars (AXPs). In addition, we should take into account that the inner core magnetic fields of magnetars can be even larger, which follows from the magnetic field flux conservation in stellar media with very large electric conductivities. The inner fields have been estimated to range from $10^{18}$ G for nuclear matter stars \cite{NS-B-3} to $10^{20}$ G for quark matter stars \cite{magnetizedfermions}. The fact that strong magnetic fields populate the vast majority of astrophysical compact objects and that they have significant consequences for several star properties has motivated many works focused on the study of the equation of state (EOS) of magnetized NS (see \cite{insignificance} and references there).
These strong magnetic fields, produced during the first instants after a collision or in the interior of NS, can create the conditions for observable QCD effects \cite{B_HIC-2}. These effects can be prominent because the magnetic fields generated are of the order of or higher than the QCD scale, $eB \gtrsim  \Lambda^2_{QCD}$.

With respect to NS, despite an extended observational and theoretical effort carried out in recent years to determine the state of the extremely high-dense nuclear-star matter existing in its interior, this still remains an open question to be answered. The physics of matter at densities beyond the nuclear density, $\rho_0=2.3\times10^{17} kg/m^3$, is a big challenge to theorists, with observations of NS being crucial for determining the correct dense-matter model. In this regard, recent very precise mass measurements of two compact objects, PSR J1614-2230 and PSR J0348+0432 with $M=1.908\pm 0.016M_{\odot}$ \cite{Demorest} and $M=2.01\pm 0.04M_{\odot}$ \cite{Antoniadis}, respectively, where $M_{\odot}$ is the solar mass, together with the analysis \cite{Miller} of the data obtained by recent NASA telescope NICER measurements of x-ray emissions from the heaviest of the precisely known $2M_{\odot}$ PSR J0740+6620, have provided some constraints on the interior composition of NS. 
These high mass values imply that the EOS of the corresponding stellar medium should be rather stiff at high densities.

Another quantity that can shed some light on the characteristics of the stellar inner matter is the speed of sound (SOS) $c_s$. This quantity is associated with how stiff the EOS is and consequently with how big the star's mass can be at a given radius. Causality imposes the absolute bound $c_s \leqslant c$ (From now on we will work in natural units with $c=1$); but its value can be even more constrained. In an isotropic medium the SOS can be obtained from 
\begin{equation}\label{SOS-Formula}
	c_s^2=\frac{\partial P}{\partial \epsilon}
\end{equation}
with $\epsilon$ the system energy density and $P$ the pressure. Then, for incompressible matter the trace of the stress-energy tensor is constant
 \begin{equation}\label{SOS-Formula-1}
	T= \epsilon+3P=constant,
\end{equation}
from which the so-called conformal limit follows $c_s^2 =1/3$.

Considering quark matter at very high densities, due to asymptotic freedom, the quarks behave as free fermions with the baryonic chemical potential as the leading parameter. Under this condition, the thermodynamic potential at zero temperature behaves as $\Omega \sim \mu^{d+1}$ \cite{Zeldovich},  where $d$ is the spatial dimension. Then, from (\ref{SOS-Formula}), the speed of sound becomes
\begin{equation}\label{SOS-Formula-3}
	c_s^2\lesssim1/d,
\end{equation}
 which for the three-dimensional space means that under such conditions the SOS in the medium reaches the conformal value $c_s^2\lesssim1/3$.

Nevertheless, at the densities reached in the core of NS this bound seems to be violated. In particular, for hybrid stars whose structure is given by an external layer of ordinary nuclear matter up to a density $\rho_c \sim 2 \rho_0$  and by an internal core beyond that density ($\rho > \rho_c$) formed by dense quark matter with a speed of sound at the conformal limit, it is impossible to reach star-mass values  $\sim 2M_{\odot}$  \cite{Bedaque}. Hence, to solve this apparent paradox it should be expected that $c_s$ must increase in the core to values significantly larger than the conformal bound \cite{Bedaque}. 

Now, the discussion of an upper limit to the SOS in NS goes back some time \cite{Zeldovich, Conformal-limit}. 
More recently, SOS larger than the conformal limit has been reported for QCD at large isospin density \cite{Beyond-Conformal-1}, in two-color QCD  \cite{Beyond-Conformal-2}
in holographic approaches  \cite{Beyond-Conformal-3}, in resumed perturbation theory \cite{Beyond-Conformal-4}, in chiral effective field theory \cite{Beyond-Conformal-5}, 
in quarkyonic matter  \cite{Beyond-Conformal-6} and in models of high-density QCD \cite{Beyond-Conformal-7}. 

In this paper, we want to add one more element into this context, that is, we will investigate the effect a strong magnetic field can have on the SOS of the different matter phases forming a hybrid star, to see if the SOS starting at values $c^2_s < 1/3$ in the hadronic phase existing in the outer core, can go beyond the conformal limit inside the quark core to finally reach the conformal limit at larger density values, as is expected in order to reach a maximum star mass $\sim 2M_{\odot}$  \cite{Bedaque}. The motivation for this investigation is that in the presence of a magnetic field the particle transverse momentum is quantized (Landau quantization) producing a discrete energy spectrum, which in the lowest Landau level (LLL) exhibits a spatial dimensional reduction to $d=1$. Thus, depending on the significance of the LLL contribution in the system EOS, we should expect from (\ref{SOS-Formula-3}) more or less a departure of the SOS from the conformal limit.

To undertake this task it is important to keep in mind that one of the consequences of the presence of a uniform magnetic field is the breaking of the system rotational symmetry, which induces new structures in the energy-momentum tensor from where the SOS is deduced. This symmetry breaking also yields an anisotropy in the EOS \cite{magnetizedfermions, PhaseTrans-B} (i.e. the system pressures along and transverse to the direction of the magnetic field are different), which has been investigated in heavy-ion collisions \cite{EoS-B-1}, as well as in astrophysics \cite{insignificance}, \cite{Laura}-\cite{EoS-B-2} and that will be the basis of our derivation.

Since, the SOS in the medium is associated with oscillations in pressure, it is expected that the anisotropy in the pressure also affects the transmission of sound along the directions aligned with and perpendicular to the field. Thus, we will start by deriving from first principles how the SOS formula given in Eq. (\ref{SOS-Formula}) is modified in the presence of a strong magnetic field. 
For this purpose, we will demonstrate that the pressure splitting is related to the anisotropy in the sound propagation by studying the first order perturbation of the thermodynamic quantities in the hydrodynamic equation (i.e. in the vanishing divergence of the corresponding stress-energy tensor). More specifically, we will show that the SOS$^2$ along the parallel and perpendicular directions to the magnetic field are equivalent to the stiffness of the parallel and perpendicular components of the EOS respectively. 

We then will analyze how a strong magnetic field can affect the SOS 
of three phases that can in principle be realized in the interior of hybrid NS. With this purpose in mind we will investigate first the SOS
of a hadronic phase that may take place in the outer core region modeled by the Walecka model \cite{Walecka}, then, the so-called magnetic dual chiral density wave (MDCDW) phase  \cite{KlimenkoPRD82}-\cite{Israel}, which is a phase that may exist at the intermediate densities  that can be realized in the inner core region close to the boundary of the hadron-quark phase transition and finally for the region closer to the star center where high densities may support a free-quark phase we consider the MIT Bag model \cite{Bag}.

The main outcome of this paper is that the SOS along these three phases is not monotonic. From values smaller than the conformal limit in the hadronic phase, it increases  in the quark-matter MDCDW phase reaching at magnetic fields $\sim 10^{18}$ G a peak with a value significantly surpassing the conformal bound at densities $\sim 3\rho_0$; while at higher densities the SOS approaches the conformal limit from below. Those are precisely the requirements pointed out in Ref. \cite{Bedaque} needed for the SOS to reach a stellar mass  $\sim 2M_{\odot}$. The inner magnetic field value $\sim 10^{18}$ G and core densities considered in this study are acceptable values for the inner core that are commonly used in the physics of NS. Nevertheless, to definitively conclude that this SOS profile in the presence of a magnetic field has a real impact in the physics of hybrid stars, a complete study of the realization of these three magnetized phases in the range of baryonic densities expected in the interior of these stars, as well as their involvement in compliance with the known physical constraints as the maximum stellar mass, are needed.

The paper is outlined as follows. In Sec. II we derive from first principles the anisotropic wavelike equation from where the longitudinal and transverse SOS in the presence of a uniform magnetic field can be extracted. We find that the anisotropic formulas relate the SOS with the stiffness of the anisotropic EOS in each corresponding direction. In Sec. III, we investigate the relevance of the LLL in the values of the SOS, showing that in the strong-field limit, when the particles are constrained to the LLL, while the transverse SOS approaches zero, the longitudinal component exceeds the conformal limit approaching the maximum value $c$. 
We then find in Sec. IV, the corresponding SOS for three possible inner density regions of hybrid stars characterized by three different phases: hadronic at low density, MDCDW phase at intermedium densities and free-quarks at higher density. From there, we analyze the entire profile for the SOS under those conditions emphasizing the role of the LLL in each phase.  In Sec. V, we summarize the paper's main results.

\section{The Speed of Sound in the presence of a magnetic field}\label{section5}

\subsection{The equations of state in a magnetic field}\label{section1}

We want to briefly introduce the anisotropic stress-energy tensor in the presence of a magnetic field from where the anisotropic EOS are obtained and from which the SOS are derived for a magnetized system.
It  is well known that the system EOS is derived from the quantum-statistical average of its stress-energy tensor. As has been stressed in Refs. \cite{magnetizedfermions, PhaseTrans-B}, the presence of a uniform magnetic field introduces a new symmetry breaking that is reflected in the covariant structures of the stress-energy tensor. Specifically, a uniform magnetic field along the third direction, $x_3$, breaks the rotational O(3) invariance making an explicit separation between the direction along the field lines and those which are transverse to them. Hence, in the presence of a medium (i.e. at finite density and/or temperature) that breaks Lorentz invariance, the magnetized medium exhibits a stress-energy tensor expanded in three independent structures
 \begin{equation}\label{Tau}
\tau^{\mu \nu}=a_1 \eta^{\mu \nu}+a_2u^\mu u^\nu +a_3\widehat{F}^{\mu \rho}\widehat{F}_{\rho}^\nu,
\end{equation}
 where $\eta^{\mu\nu}$ is the Minkowski metric, $u_\mu$ is the medium four-velocity, which in the rest frame takes the value $u_\mu =(1,\overrightarrow{0})$,  and $\widehat{F}^{\mu \rho}=F^{\mu \rho} /B$ is the normalized electromagnetic strength tensor.
 
The scalar coefficients $a_i$, $i=1,2,3$ are found from the quantum-statistical average of the energy-momentum tensor, as
 \begin{equation}\label{Tau2}
\frac{1}{\beta V}\langle\hat{\tau}^{\mu \nu}\rangle=\Omega \eta^{\mu \nu}+(\mu \rho+TS)u^\mu u^\nu +BM \eta_{\perp}^{\mu \nu}
\end{equation}
They are given in terms of the system thermodynamic potential $\Omega$ , the average particle-number density, $\rho=-(\partial \Omega / \partial \mu)_{V,T}$, the entropy, $S=-(\partial \Omega / \partial T)_{V, \mu}$, and the system magnetization $M=-(\partial \Omega / \partial B)_{V,T,\mu}$. In (\ref{Tau2}), $V$ is the system volume, $B$ is the magnetic field, $\mu$ is the chemical potential, $\beta=1/T$ is the inverse absolute temperature and $\eta_{\perp}^{\mu \nu}=diag(0,-1,-1,0)$. 

From (\ref{Tau2}), we can find the system anisotropic EOS by calculating the different components of $\big<\hat{\tau}^{\mu\nu} \big>$, which are given in terms of the energy density $ \varepsilon$, parallel pressure $P_\parallel$ and  perpendicular  pressure $P_\perp$ as \cite{magnetizedfermions}
\begin{equation} \label{energy-pressure}
\epsilon=\frac{1}{{\beta}V}\big<\hat{\tau}^{00}\big>,\quad P_\parallel=\frac{1}{\beta V}\big<\hat{\tau}^{33}\big>,\quad P_\perp=\frac{1}{{\beta}V}\big<\hat{\tau}^{\perp\perp}\big>
\end{equation}
As is known, for NS physics, since $\mu \gg T$, the $T=0$ limit is a justified approximation and the EOS reduces to
\begin{equation} \label{energy-density-1}
    \epsilon=\Omega+\mu \rho+\frac{B^2}{2},
\end{equation}
\begin{equation} \label{Pressures-EOS-1}
    P_\parallel=-\Omega-\frac{B^2}{2},\quad{P_\perp=-\Omega-MB+\frac{B^2}{2}}.
\end{equation}
In (\ref{energy-density-1})-(\ref{Pressures-EOS-1}) the quadratic terms in $B$ arise from the Maxwell contribution to the energy momentum tensor 
\begin{equation}\label{maxwellset}
	\tau^{\mu\nu}_M=\frac{B^2}{2}\left(\eta^{\mu\nu}_{\parallel}-\eta^{\mu\nu}_{\perp}\right).
\end{equation}

Since the EOS is the basis for the calculation of the SOS, the EOS anisotropy will be transferred to it as we will see as follows.

\subsection{The Speed of Sound with Anisotropic Pressures}\label{subsection4a}

In an isotropic system the speed of sound squared (SOS$^2$) is known to be given by the differential slope of the P vs $\varepsilon$ curve,  $\partial{P}/\partial{\varepsilon}$. 
Due to the existence of a pressure anisotropy in the presence of a magnetic field, a corresponding anisotropy in the speed of sound is expected to exist. As follows we will show that the speed of sound becomes anisotropic and given by
\begin{equation}\label{SOS}
	c_{\parallel}^2=\Big[\frac{\partial{P_{\parallel}}}{\partial{\epsilon}}\Big]_{B},\:\:\:\:\:\:\:\:\:\:c_{\perp}^2=\Big[\frac{\partial{P_{\perp}}}{\partial{\epsilon}}\Big]_{B}\end{equation}
where $c_{\parallel}$ and $c_{\perp}$ are the sound velocities along and perpendicular to the field direction, respectively.

Our goal now is to justify (\ref{SOS}) from general principles. From the results (\ref{Tau2}), (\ref{energy-density-1}) and (\ref{Pressures-EOS-1}) we can express the quantum-statistical average of the stress-energy tensor as

\begin{equation}\label{stressenergy-2}
	\frac{1}{\beta V}\langle\hat{\tau}^{\mu \nu}\rangle=-P_{\parallel}\eta^{\mu\nu}+\left(\varepsilon+P_{\parallel}\right)u^{\mu}u^{\nu}+\left(P_{\parallel}-P_{\perp}\right)\eta^{\mu\nu}_{\perp}.
\end{equation}

In (\ref{stressenergy-2}), $\varepsilon$ is the system energy density in a frame comoving with the fluid. In a general local Minkowski frame, where the fluid is moving with four velocity $u_\mu$, the energy density is given by
\begin{equation}\label{E}
	\epsilon=-P_{\parallel}+\left(\varepsilon+P_{\parallel}\right)\gamma^2.
\end{equation}
where $\gamma=[1-v^2]^{-1/2}$ is the Lorentz factor and $v$ is the magnitude of the system spatial velocity in the  local Minkowski frame.

Assuming there are no external four forces, the hydrodynamics of the system is determined in the local Lorentz frame by the vanishing divergence of the stress-energy tensor, which is associated with the conservation of energy and linear momentum,
\begin{equation}\label{divergence}
   \partial_{\nu} \langle\hat{\tau}^{\mu \nu}\rangle=0
\end{equation}

Substituting (\ref{stressenergy-2}) into (\ref{divergence}) and using (\ref{E}) to simplify, (\ref{divergence}) can be broken into components as follows
\begin{equation}\label{0comp}
	\partial_0{\epsilon}+\nabla\cdot\left[(\epsilon+P_{\parallel})\vec{v}\right]=0,\quad\quad\mu=0,
\end{equation}

\begin{equation}\label{12comp}
\begin{aligned}
	&\partial_{\mu}P_{\perp}+\partial_0\left[(\epsilon+P_{\parallel})v^{\mu}\right]+\partial_j\left[(\epsilon+P_{\parallel})v^{\mu}v^j\right]=0,\quad\quad\mu=1,2,
\end{aligned}
\end{equation}

\begin{equation}\label{3comp}
\begin{aligned}
&\partial_3P_{\parallel}+\partial_0\left[(\epsilon+P_{\parallel})v^3\right]+\partial_j\left[(\epsilon+P_{\parallel})v^3v^j\right]=0,\quad\quad\mu=3
\end{aligned}
\end{equation}
where we introduced the spatial index $j=1,2,3$. Using (\ref{0comp}) to eliminate $\partial_0{\epsilon}$ from (\ref{12comp}) and (\ref{3comp}) we arrive at
\begin{equation}\label{12comp-2}
	(\epsilon+P_{\parallel})\left[\partial_0v^{\mu}+v^j\partial_jv^{\mu}\right]=-\partial_{\mu}P_{\perp}-v^{\mu}\partial_0P_{\parallel},\:\:\:\:\:\:\mu=1,2
\end{equation}

\begin{equation}\label{3comp-2}
	(\epsilon+P_{\parallel})\left[\partial_0v^3+v^j\partial_jv^3\right]=-\partial_3P_{\parallel}-v^3\partial_0P_{\parallel},\:\:\:\:\:\:\mu=3.
\end{equation}

Now we consider a system that is at rest relative to a local Minkowski frame (i.e. with $v=0$) and where $\epsilon$, $B$, $P_{\parallel}$, and $P_{\perp}$ are uniform and constant throughout. Further, let the system be perturbed, ${\epsilon}\to\epsilon+\delta\epsilon$,  $P_{\parallel}\to{P_{\parallel}+\delta{P_{\parallel}}}$, $P_{\perp}\to{P_{\perp}+\delta{P_{\perp}}}$, and in such a way that the fluid velocity in the local frame is also perturbed as
$\vec{v}\to\delta{\vec{v}}$. Substituting these changes into (\ref{0comp}), (\ref{12comp-2}), and (\ref{3comp-2}) and keeping terms to first order in the perturbation we get

\begin{equation}\label{0comp-3}
    \partial_0{\delta\epsilon}+(\epsilon+P_{\parallel})\partial_j(\delta{v^j})=0,\:\:\:\:\mu=0,
\end{equation}
and 
\begin{equation}\label{12comp-3}
    (\epsilon+P_{\parallel})\partial_0\delta{v^{\mu}}=-\partial_{\mu}{\delta{P_{\perp}}},\:\:\:\:\:\:\:\mu=1,2,
\end{equation}

\begin{equation}\label{3comp-3}
    (\epsilon+P_{\parallel})\partial_0\delta{v^3}=-\partial_{3}{\delta{P_{\parallel}}},\:\:\:\:\:\:\mu=i=3.
\end{equation}

Taking the time derivative of (\ref{0comp-3}), the three divergence of (\ref{12comp-3}) and (\ref{3comp-3}), and combining the results we arrive at 
\begin{equation}\label{wave-1}
    \frac{\partial^2\delta{\epsilon}}{\partial{t}^2}-\left( \frac{\partial^2}{\partial{x}^2}+ \frac{\partial^2}{\partial{y}^2}\right)\delta{P_{\perp}}-\frac{\partial^2}{\partial{z}^2}\delta{P_{\parallel}}=0.
\end{equation}

Considering that $\epsilon$, $P_{\parallel}$, and  $P_{\perp}$ are in general functions of $\mu$ and $B$, the variations in $\epsilon$, $P_{\parallel}$, and $P_\perp$ are given by 
\begin{equation}\
\begin{aligned}
    \delta{\epsilon}&=\frac{\partial{\epsilon}}{\partial{\mu}}\delta{\mu}+\frac{\partial{\epsilon}}{\partial{B}}\delta{B},
\\
 \delta{P_{\parallel}}&=\frac{\partial{P_{\parallel}}}{\partial{\mu}}\delta{\mu}+\frac{\partial{P_{\parallel}}}{\partial{B}}\delta{B},
\\
\delta{P_{\perp}}&=\frac{\partial{P_{\perp}}}{\partial{\mu}}\delta{\mu}+\frac{\partial{P_{\perp}}}{\partial{B}}\delta{B}.
\end{aligned}
\end{equation}

In order to consider the response only to mechanical perturbations we take $\delta{B}=0$ (taking $\delta{B}\neq{0}$ would amount to a perturbation in the electromagnetic field), then (\ref{wave-1}) becomes
\begin{equation}\label{mu-Eq}
   \frac{\partial{\epsilon}}{\partial{\mu}} \frac{\partial^2\delta{\mu}}{\partial{t}^2}-\frac{\partial{P_{\perp}}}{\partial{\mu}}\left( \frac{\partial^2}{\partial{x}^2}+ \frac{\partial^2}{\partial{y}^2}\right)\delta{\mu}-\frac{\partial{P_{\parallel}}}{\partial{\mu}}\frac{\partial^2\delta{\mu}}{\partial{z}^2}=0,
\end{equation}
which can be written as
\begin{equation}\label{Wave-Eq}
    \frac{\partial^2\delta{\mu}}{\partial{t}^2}-(c_s^{\perp})^2\left( \frac{\partial^2}{\partial{x}^2}+ \frac{\partial^2}{\partial{y}^2}\right)\delta{\mu}-(c_s^{\parallel})^2\frac{\partial^2}{\partial{z}^2}\delta{\mu}=0,
\end{equation}
after we identify the sound velocity components
\begin{equation}\label{velocidades}
 (c_s^{\perp})^2=\frac{\partial{P_{\perp}}}{\partial{\mu}}/ \frac{\partial{\epsilon}}{\partial{\mu}}=\Big[\frac{\partial{P_{\perp}}}{\partial{\epsilon}}\Big]_{B},\:\:(c_s^{\parallel})^2=\frac{\partial{P_{\parallel}}}{\partial{\mu}}/ \frac{\partial{\epsilon}}{\partial{\mu}}=\Big[\frac{\partial{P_{\parallel}}}{\partial{\epsilon}}\Big]_{B}
\end{equation}

Then, from (\ref{Wave-Eq})-(\ref{velocidades}), we see that $c_s^{\parallel}$ and $c_s^{\perp}$ describe in the  local Minkowski frame the SOS along and transverse to the magnetic field direction respectively.
In Section IV, we calculate the SOS for three different magnetized phases that can be realized inside a hybrid star as the density decreases from the outer to the inner core.

\section{The role of the lowest Landau level in the speed of sound}

As we discussed in the Introduction, a charged fermion system in the LLL is confined to one spatial dimension and from Eq. (\ref{SOS-Formula-3}) it is expected that its SOS is significantly beyond the conformal limit.  In this section, we will calculate the SOS for a generic charged fermion system in a sufficiently strong magnetic field able to confine the particles into the LLL.

Thus, we start from the zero temperature many-particle thermodynamic potential for a charged fermion species (indexed by "i"), which is known to be (see for instance \cite{PhaseTrans-B}) 
\begin{equation}\label{Omega-f}
	\Omega_i=-g_ie_iB\int_{-\infty}^{\infty}\frac{dp_3}{\left(2\pi\right)^2}\left\{\left(\mu_i-E_i\left(l=0\right)\right)\Theta\left(\mu_i-E_i\left(l=0\right)\right)+2\displaystyle\sum_{l=1}^{\infty}\left(\mu_i-E_i\right)\Theta\left(\mu_i-E_i\right)\right\},
\end{equation}
where the energy spectrum is given by
\begin{equation}\label{energy-spectrum}
	E_i=\sqrt{p_3^2+2e_iBl+m_i^2}, \quad l=0,1,2,\cdots
\end{equation}
with $l$ denoting the Landau level numbers and $\mu_i$ being the chemical potential of the fermion species $i$, $g_i$ is the spin degeneracy given by $g_i=2-\delta_{l0}$ and $e_i$ being the corresponding fermion charge. Notice that for $l=0$ the energy spectrum obtained from (\ref{energy-spectrum}) corresponds to a particle moving in one spatial dimension.  

The Heaviside functions in (\ref{Omega-f}) induce cutoffs in both the $p_3$ integral and the sum over $l$. Hence, performing the momentum integrals we get 

\begin{equation}
\begin{aligned}
	\Omega_i=&-\frac{g_ie_iB}{\left(2\pi\right)^2}\Bigg[\mu_i\sqrt{\mu_i^2-m_i^2}-m_i^2\ln\left\lvert\frac{\mu_i+\sqrt{\mu_i^2-m_i^2}}{m_i}\right\rvert
\\
&+2\displaystyle\sum_{l=1}^{l_{max}}\left\{\mu_i\sqrt{\mu_i^2-\left(2e_iBl+m_i^2\right)}-\left(2e_iBl+m_i^2\right)\ln\left\lvert\frac{\mu_i+\sqrt{\mu_i^2-\left(2e_iBl+m_i^2\right)}}{\sqrt{2e_iBl+m_i^2}}\right\rvert\right\}\Bigg],
\end{aligned}
\end{equation}

where 
\begin{equation}\label{l-max}
	l^{max}_i=\left\lfloor{x_i}\right\rfloor,\quad x_i=\frac{\mu_i^2-m_i^2}{2e_iB}.
\end{equation}

From this last relation we see that when $2e_iB>\mu_i^2$ the sum over Landau levels in $\Omega_i$ collapses leaving only the LLL term. For magnetic fields satisfying this condition we have 

\begin{equation}\label{Omega-LLL}
	\Omega_i^{LLL}=-\frac{g_ie_iB}{\left(2\pi\right)^2}\left[\mu_i\sqrt{\mu_i^2-m_i^2}-m_i^2\ln\left\lvert\frac{\mu_i+\sqrt{\mu_i^2-m_i^2}}{m_i}\right\rvert\right].
\end{equation}

Therefore, from Eqs. (\ref{energy-density-1})-(\ref{Pressures-EOS-1}) and (\ref{Omega-LLL}) we have 
\begin{equation}\label{EOS-1}
	P_{\parallel}^{LLL}=-\Omega_i^{LLL}-B^2/2=\frac{g_ie_iB}{\left(2\pi\right)^2}\left[\mu_i\sqrt{\mu_i^2-m_i^2}-m_i^2\ln\left\lvert\frac{\mu_i+\sqrt{\mu_i^2-m_i^2}}{m_i}\right\rvert\right]-B^2/2,
\end{equation}

\begin{equation}\label{EOS-2}
	P_{\perp}=-\Omega_i^{LLL}+B \frac{\partial \Omega_i^{LLL}}{\partial B}+B^2/2=B^2/2,
\end{equation}

\begin{equation}\label{EOS-3}
	\epsilon=\Omega_i^{LLL}-\mu_i \frac{\partial \Omega_i^{LLL}}{\partial \mu_i}+B^2/2=\frac{g_ie_iB}{\left(2\pi\right)^2}\left[\mu_i\sqrt{\mu_i^2+m_i^2}-m_i^2\ln\left\lvert\frac{\mu_i+\sqrt{\mu_i^2+m_i^2}}{m_i}\right\rvert\right]+B^2/2,
\end{equation}

The result in (\ref{EOS-2}) is due to the fact that the thermodynamic potential in the LLL (\ref{Omega-LLL}) is proportional to $B$, thus we have that $BM=-\Omega_i^{LLL}$. 

 Now, using Eq.  (\ref{velocidades}) the SOS$^2$ parallel and perpendicular components are given in the  $m_i \ll \mu_i$ limit by
\begin{equation}\label{SOSQParLLL}
	(c_S^{\parallel})^2=1- \frac{m_i^2}{\mu_i^2}\simeq 1, \quad\quad  (c_s^{\perp})^2=0
\end{equation}

Hence, for the LLL to play a significant  role in the system and specifically for it to exhibit a SOS beyond the conformal limit, the following parameter  order should exist: $m^2_i < \mu^2_i \lesssim  B$. In the next section, we will calculate the SOS for three possible inner NS phases: A hadronic phase and two quark phases. In particular, we will show, that quark matter at intermediate densities and in the presence of a sufficiently strong magnetic field ($\sim 10^{18}$ G)  can reach an SOS with a value larger than the conformal bound.

\section{The SOS profile inside a magnetized hybrid Neutron Star}\label{subsection4}
In this section we investigate how the value of the SOS changes in the presence of a magnetic field with changing energy density as we move inside a hybrid star from the region closed to the outside of the core to the star center. In particular, the role played in these results by the LLL will be emphasized.  

To find the SOS of the hadronic and quark phases that can be realized in the different star regions,  we rely on their corresponding  EOS in the presence of a magnetic field. The role of the magnetic field in the SOS depends on its relative value with respect to the other  parameters in play. In the case of hadrons, to have a significant effect associated with the LLL, a field of $\sim 10^{20}$ G would be needed \cite{insignificance} in order to satisfy the relation $B > m_p^2$, where $m_p$ is the proton mass. But such a field will make the NS unstable by making the inner parallel pressure extremely small \cite{Laura}. Thus, in the case of the hadronic system it is legitimate to use the weak-field approximation (WFA), because even a field $\sim 10^{18}$ G is several orders smaller than the square of the proton mass. In this case a large number of LL's will equally contribute.

For the inner core, where the quark phase can be realized, it is more expected that the baryon density goes from intermediate densities to high densities as we approach the stellar center. At intermediate densities, it has long been argued that at relatively low temperatures, phases with spatially inhomogeneous chiral condensates will be favored over the homogeneous ones. Such spatially inhomogeneous phases have been found in the large-N limit of QCD \cite{largeNQCD,largeN}, in NJL models \cite{DCDW}-\cite{PRD85-074002}, and in quarkyonic matter \cite{q-chiralspirals}-\cite{ferrer-incera-sanchez}. In all the cases, single-modulated chiral condensates are energetically favored over chiral condensates with higher-dimensional modulations.

Notwithstanding, single-modulated phases in three spatial dimensions are known to be unstable against thermal fluctuations at any arbitrarily small finite temperature, a phenomenon known in the literature as Landau-Peierls instability \cite{Landau-Peirls Inst}. 
This instability signals the lack of long-range correlations at any finite temperature and hence the lack of a true order parameter. A solution to this problem was found in \cite{Ferrer-Incera19} by considering the Dual-Chiral-Density-Wave phase in the presence of a magnetic field \cite{KlimenkoPRD82}-\cite{Israel}, which was then called the MDCDW phase (M for magnetic). In \cite{Ferrer-Incera19} it was shown that a background magnetic field introduces new structures in the system that are consistent with the symmetry group that remains after the explicit breaking of the rotational and isospin symmetries by the magnetic field and thus removing the Landau-Peierls instability. This is so, because the new terms not only modify the condensate minimum equations, but they also lead to a linear, anisotropic spectrum of the thermal fluctuations, which lacks soft transverse modes. Soft transverse modes are the essence of the Landau-Peierls instability because they produce infrared divergences in the mean square of the fluctuation field that in turn wipe out the average of the condensate at any low temperature. 

In addition to that, the MDCDW phase exhibits three other properties that are important for the astrophysics of NS: First, the temperature needed to evaporate the inhomogeneous condensate for fields $\sim 10^{18}$ G is higher than the stellar temperatures for the whole range of densities characteristic for NS \cite{Will}, second, when considering the TOV equations, it was proved in \cite{InhStars} that the maximum stellar mass of a hybrid star with a quark-matter core in this phase satisfies the maximum mass observation constraints ($M \gtrsim  2M_{\odot}$) \cite{Demorest, Antoniadis}, and third, in \cite{PRD20} it was shown that if the NS core is formed by quarks in the MDCDW phase the heat capacity will be well above the lower limit expected for NS ($C_V\gtrsim 10^{36}(T/10^8)$ erg/K) \cite{Cv-NS}. This limit was established by
long-term observations of NS temperatures in the range from months to years after accretion outburst together with continued observations on timescales of years.
On the other hand, this lower-limit value  put out of the game the matter components that exhibit superfluidity or superconductivity of any kind, since as showed in \cite{PRD20} all these cases are exponentially damped. This will lead to the striking conclusion that, if the only quark matter state to be realized in NS interior is the color superconducting CFL/MCFL phases \cite{CFL, Lec-Notes}, quarks will be ruled out from forming part of the NS core \cite {Cv-NS}. 
In view of all these achievements of the MDCDW phase as a viable candidate to be considered as the possible quark phase to be realized in the intermediate density region of the inner core of NS, we will investigate in this section its corresponding SOS to show that at sufficiently high magnetic fields it can exhibit a peak with values larger than the conformal limit $\sqrt{1/3}$.

Finally, at asymptotically high baryon densities where $\mu^2 > B, m_q^2$, we will consider a free-quark model in a relatively weak magnetic field to show that at those densities the SOS can reach from below the conformal limit as it was expected in Ref. \cite{Bedaque}. In this region the role of the magnetic field is not significant, but the one of the large baryonic chemical potential.

\subsection{ The SOS at low densities: The Magnetized Hadronic Phase}\label{section2-1} 

In this section, for the sake of completeness and to facilitate the reader's understanding, we summarize the results for the anisotropic EOS of the hadronic phase in the presence of a magnetic field. 
The EOS of hadronic matter in the presence of a magnetic field were originally calculated in Refs. \cite{EOS-Hadrons}.  In those works the pressure anisotropy was not considered. Latter on, after  \cite{magnetizedfermions}, the EOS anisotropy was taken into account and applied to a variety of phenomena in different contexts (see \cite{insignificance}, \cite{PhaseTrans-B}-\cite{EoS-B-2},  \cite{neutrons}-\cite{EOS-Anisotropic} and references therein). In what follows, we summarize the results for the corresponding energy density, as well as for the longitudinal and transverse pressures of the hadronic phase described by the non-linear Walecka (NLW) model \cite{Walecka}. 

We consider a system with baryon content comprised only of neutrons and protons ($b=n,p$) and lepton content comprised only of electrons ($l=e$). The meson fields will be considered in the mean-field approximation (MFA), where only the expectation values $\bar{\sigma}=\left<\sigma\right>$, $\bar{\omega}_0=\left<\omega_0\right>$, $\bar{\rho}_0=\left<\rho^0\right>$ contribute into the one-loop thermodynamic potential. 

The EOS for this system has been calculated in detail in Ref. \cite{PhaseTrans-B}. Thus, we are only giving the results for the energy density, perpendicular and parallel pressures respectively,
\begin{equation}\label{PressuresHS}
	\begin{aligned}
	&\varepsilon={\Omega_f}^{H}+\displaystyle\sum_{i}E^F_i{\rho}_i^{{H}}+\frac{B^2}{2}+\left\{\frac{1}{2}m_{\sigma}^2\bar{\sigma}^2+U(\bar{\sigma})+\frac{1}{2}m_{\omega}^2{\bar{\omega}_0}^2+\frac{1}{2}m_{\rho}^2{\bar{\rho}_0}^2\right\},
\\
&P_{\perp}=-{\Omega_f}^{H}-BM_f^{H}+\frac{B^2}{2}+\left\{-\frac{1}{2}m_{\sigma}^2\bar{\sigma}^2-U(\bar{\sigma})+\frac{1}{2}m_{\omega}^2{\bar{\omega}_0}^2+\frac{1}{2}m_{\rho}^2{\bar{\rho}_0}^2\right\},
\\
&P_{\parallel}=-{\Omega_f}^{H}-\frac{B^2}{2}+\left\{-\frac{1}{2}m_{\sigma}^2\bar{\sigma}^2-U(\bar{\sigma})+\frac{1}{2}m_{\omega}^2{\bar{\omega}_0}^2+\frac{1}{2}m_{\rho}^2{\bar{\rho}_0}^2\right\},
	\end{aligned}
\end{equation}
Here,
\begin{equation}\label{Potential}
	U(\sigma)=\frac{1}{3}cm_n(g_{\sigma{N}}\sigma)^3+\frac{1}{4}d(g_{\sigma{N}}\sigma)^4.
\end{equation}
is the  scalar self-interaction potential (see Table 1 for a list of parameter values), and
\begin{equation}
	\Omega_f^H=\Omega_n+\Omega_p+\Omega_e.
\end{equation}
is the sum of the neutron, proton and electron one-loop thermodynamic potentials given  respectively in the WFA by
\begin{eqnarray} \label{ThermoPotentialFiniteDensity}
\Omega_n&=&\frac{-1}{48\pi^2}\Bigg\{2\left(\sqrt{1-\left(\frac{m^*_n+k_nB}{E^F_n}\right)^2}+\left(B\to-B\right)\right){E^F_n}^4\nonumber
\\
&+&4k_nB\left(\sin^{-1}\left[\frac{m^*_n+k_nB}{E^4_n}\right]-\left(B\to-B\right)\right){E^F_n}^3 \nonumber
\\
&+&\left[\left(m^*_n+k_nB\right)^3\left(3m^*_n-k_nB\right)\left(\ln\left[1+\sqrt{1-\frac{m^*_n+k_nB}{E^F_n}}\right]-\ln\left\lvert\frac{m_n^*+k_nB}{E^F_n}\right\rvert\right)+\left(B\to-B\right)\right] \nonumber
\\
&+&\left[\left(8k_nB\left(m^*_n+k_nB\right)-5\left(m_n^*+k_nB\right)^2\right)\sqrt{1-\frac{m^*_n+k_nB}{E^F_n}}+\left(B\to-B\right)\right]\Bigg\},
\end{eqnarray}
\begin{equation}
\Omega_p= \frac{-1}{24\pi^2}\left\{\left(2{E^F_p}^4-5{m_p^*}^2{E_p^F}^2\right)\sqrt{1-\left(\frac{m_p^*}{E_p^F}\right)^2}+\left(3{m_p^*}^4+2\left(eB\right)^2\right)\ln\left[\frac{{E_p^F}+\sqrt{{E_p^F}^2-{m_p^*}^2}}{{m_p^*}}\right]\right\},
\end{equation}
\begin{equation} \label{Omega-e}
\Omega_e=\frac{-1}{24\pi^2}\left\{\left(2{E_e^F}^4-5{m_e}^2{E_e^F}^2\right)\sqrt{1-\left(\frac{m_e}{E_e^F}\right)^2}+\left(3{m_e}^4+2\left(eB\right)^2\right)\ln\left[\frac{{E_e^F}+\sqrt{{E_e^F}^2-{m_e}^2}}{{m_e}}\right]\right\},
\end{equation}
with $m_n^*=m_n-g_{\sigma{N}}\bar{\sigma}$, $m_p^*=m_p-g_{\sigma{N}}\bar{\sigma}$.
\begin{table}[ht]
\caption{Saturation density $\rho_0$, scalar, $g_{\sigma{N}}$, and vector meson-nucleon, $g_{\omega{N}}$, $g_{\rho{N}}$, couplings as well as meson self interaction coefficients, $c$, $d$,  chosen to reproduce the binding energy, baryon density, symmetry energy coefficient and effective mass at nuclear saturation for a compression modulus $K=300$MeV, as reported in Ref. \cite{Rabhi, Table}.}
\centering
  \begin{tabular}{ | m{1cm}| m{1cm} | m{1cm} |m{1.5cm} |m{1.5cm} | m{1.5cm}|} 
    \hline
 $g_{\sigma{N}}$ & $g_{\omega{N}}$  & $g_{\rho{N}}$ & c & d&$\rho_0$  \\ 
\hline
 8.910 & 10.610 & 8.196 & 0.002947 & -0.001070 & 0.153fm$^{-3}$\\ 
  &  & & & & \\ 
\hline
  \end{tabular}
  \label{parameters}
\end{table}

In (\ref{PressuresHS}), $\rho_i^{H}=-\partial{\Omega_f}^{H}/\partial{\mu_i}$ are the $i^{th}$ particle species number densities at zero temperature and finite density, which are given by
\begin{equation}\label{numberdensitiesHS}
\begin{aligned}
	{\rho}_n&=\frac{1}{3\pi^2}\left({E^F_n}^2-{m^*_n}^2\right)^{\frac{3}{2}},
\\
	{\rho}_p&=\frac{1}{3\pi^2}\left[\left({E^F_p}^2-{m^*_p}^2\right)^{\frac{3}{2}}+\frac{(eB)^2}{4\sqrt{{E^F_p}^2-{m^*_p}^2}}\right],
\\
	{\rho}_e&=\frac{1}{3\pi^2}\left[\left({E^F_e}^2-m_e^2\right)^{\frac{3}{2}}+\frac{(eB)^2}{4\sqrt{{E^F_e}^2-m_e^2}}\right],
\end{aligned}
\end{equation}
$E_i^F$ are the Fermi energies
\begin{equation} \label{Chem-Pot}
	E_n^F=\mu_n-g_{\omega{N}}{\bar{\omega}^0}-g_{\rho{N}}\tau_{3_n}\bar{\rho}^0,  \quad E_p^F=\mu_p-g_{\omega{N}}\bar{\omega}^0-g_{\rho{N}}\tau_{3_p}\bar{\rho}^0,\quad E_e^F=\mu_e,
\end{equation}
and $M_f^H=-\partial \Omega_f^H/\partial B$ is the system magnetization. In (\ref{numberdensitiesHS}) we neglected  into $\rho_n$ the contribution of the magnetic-field/neutron-anomalous-magnetic-moment (B-N-AMM) interaction.

We will work under the assumption that the system is beta equilibrated 
\begin{equation}\label{hadronchempotentials}
	\mu_n=\mu, \quad \mu_p=\mu_n-\mu_e.
\end{equation}
as well as neutral
\begin{equation}\label{H-neutrality}
	\frac{\partial \Omega_f^H}{\partial \mu_e}=\rho_p-\rho_e=0
\end{equation}
while taking into account the condensate solutions to the minimum equations
\begin{equation}\label{condensates}
\frac{\partial{\Omega^H}}{\partial{\bar{\sigma}}}=\frac{\partial{\Omega^H}}{\partial{\bar{\omega}_0}}=\frac{\partial{\Omega^H}}{\partial{\bar{\rho}_0}}=0,
\end{equation}
where 
	\begin{equation}
		\Omega^H=\Omega^H_f+\frac{B^2}{2}+\left\{\frac{1}{2}m_{\sigma}^2\bar{\sigma}^2+U(\bar{\sigma})-\frac{1}{2}m_{\omega}^2{\bar{\omega}_0}^2-\frac{1}{2}m_{\rho}^2{\bar{\rho}_0}^2\right\}
	\end{equation}
is the full thermodynamic potential.
 Thus, in the hadronic system there are six independent parameters $B$, $\mu$, $\mu_e$, $\bar{\sigma}$, $\bar{\omega}_0$, and $\bar{\rho}_0$. Taking into account the neutrality condition (\ref{H-neutrality}) and the condensate equations (\ref{condensates}), the number of independent parameters reduces to two. Finally, fixing $B$,  the EOS becomes an implicit function of $\mu$, which is the baryonic chemical potential.
As we already pointed out, in the approach under consideration we are working in the WFA, where the Landau sums were approximated using the Euler-Maclaurin formula (see Appendix A in \cite{insignificance}) with order parameter $(eB)$.

In \cite{insignificance}, it was found that for magnetized charged fermions, the interaction of the magnetic field with the anomalous magnetic moment (B-AMM) does not significantly affect the systems EOS, so we only include here the B-N-AMM interaction for neutrons via the coupling $k_n=\mu_Ng_N/2$, where $\mu_N=|e|\hbar/2m_p=3.15\times10^{-18}$MeV$/$G is the nuclear magneton \cite{Lande-Factor}, $m_p$ is the proton mass and $g_N$ is the Land\'{e} $g$ factor for the neutron.

\begin{figure}
\begin{center}
\begin{tabular}{ccc}
  \includegraphics[width=7cm]{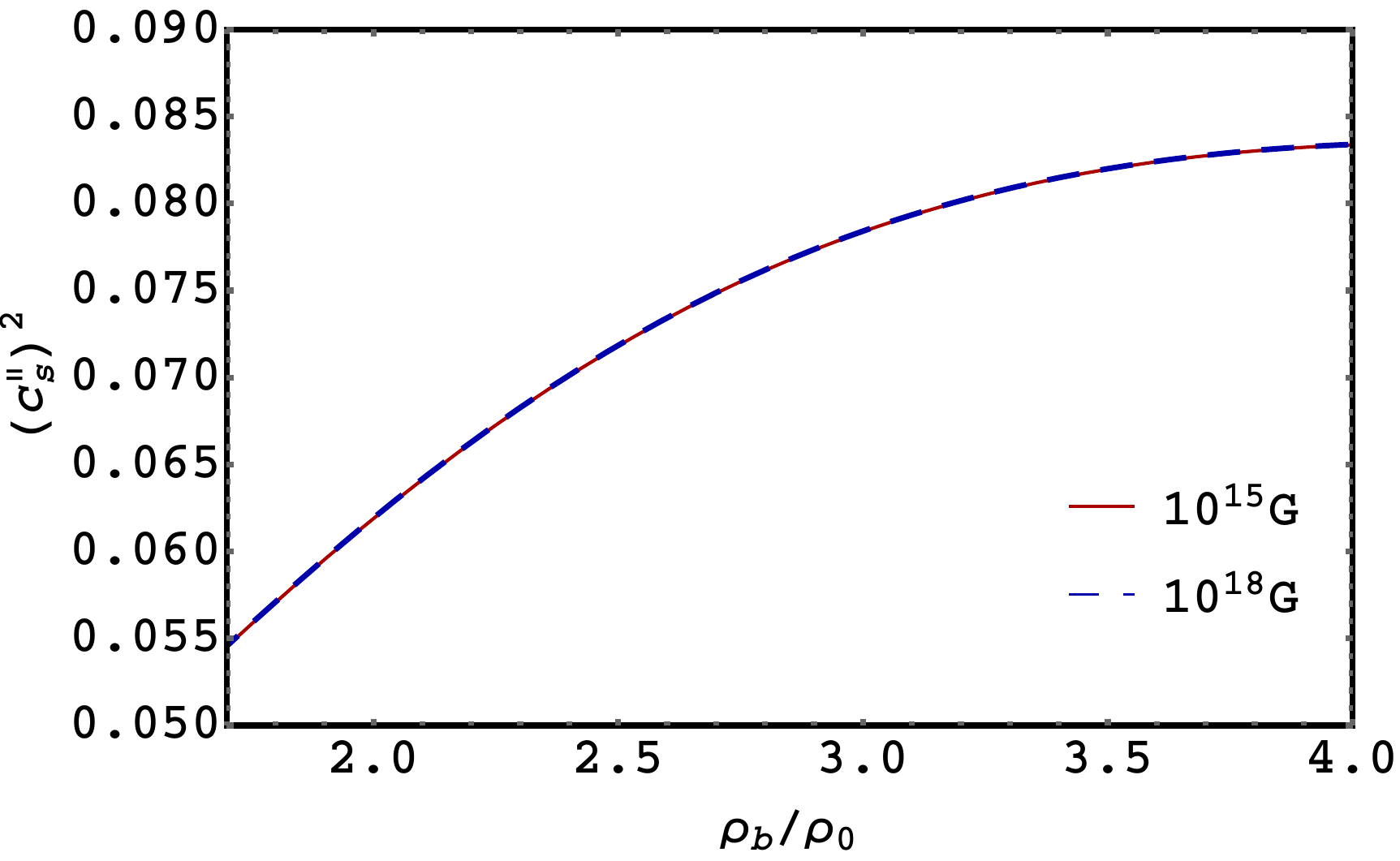} & \includegraphics[width=7cm]{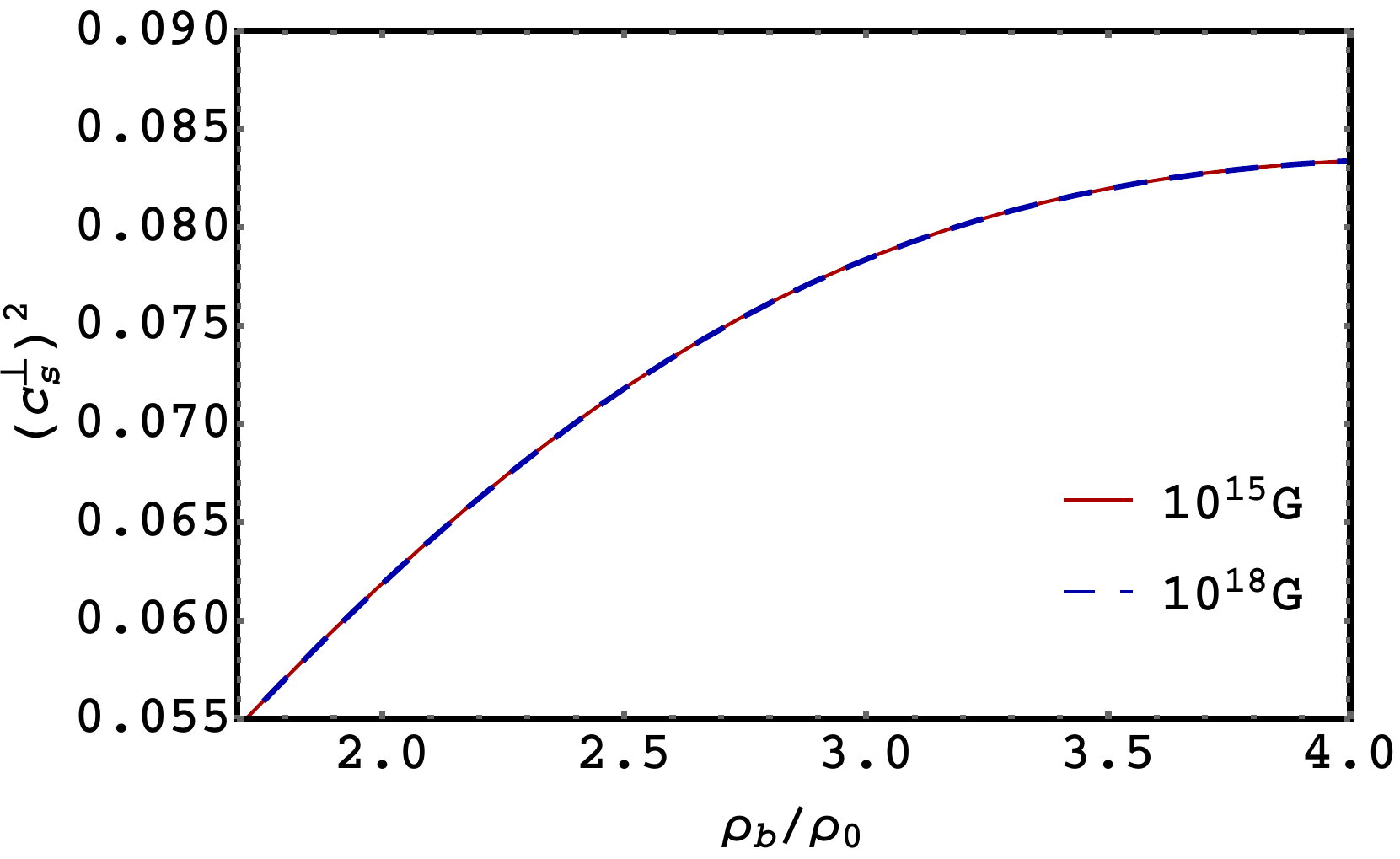}\\
(a) & (b)  \\
  \\
  \end{tabular}
    \end{center}
    \caption{(Color online) Parallel (a) and Perpendicular (b) SOS$^2$ in the hadronic system versus normalized baryonic density $\rho_b/\rho_0$ ($\rho_0$ is the nuclear saturation density)  at different magnetic field values without the contribution of the magnetic field interaction with the neutron anomalous magnetic moment.}
     \label{SOS-Hadrons}
\end{figure}

 To find the SOS for the hadronic phase we rely on the  EOS relations  (\ref{PressuresHS}) and use the general Eqs. (\ref{velocidades})-(\ref{SOS}). 
 In Fig. \ref{SOS-Hadrons} the parallel and perpendicular SOS$^2$ as defined in (\ref{velocidades}) are plotted at different values of the magnetic field. There, we have plotted the SOS$^2$ versus the normalized baryonic density in units of the nuclear saturation density $\rho_0=0.153$ fm$^{-3}$. The considered magnetic field values are concordant with the WFA for those densities.

 From Fig. \ref{SOS-Hadrons} we observe that the values of the SOS increase with the baryon density. This is a consequence of the competition between the fluid stiffness, which increases with density and the density itself which is expressed by the general Newton-Laplace formula $c_s=\sqrt{K_s/\rho}$, where $K_s$ is the coefficient of stiffness (also called the elastic bulk modulus) and $\rho$ the fluid density.  That is, the increase in density produces an increase in $K_s$, which is larger than the effect of the medium's inertial increase measured by  $\rho$ itself in the Newton-Laplace formula. It is also possible to give a second reading to the observed behavior of the SOS$^2$ with $\rho$ in Fig. \ref{SOS-Hadrons}. The graphs are also indicating that the medium increases its resistance to deformation with density, thus allowing the pressure perturbation to travel quicker without significant losses produced by the medium's deformation. The physical scenario we are describing here for dense media is framed by the same principle that explains why the speed of sound is faster in solids than in liquids or gasses. 
 
On the other hand, from the graphs in Fig. \ref{SOS-Hadrons} it can be seen that they do not display a significant splitting between $c_\|$  and  $c_\bot$. The fact that even at fields of $\sim 10^{18}$ G the SOS of the magnetized hadronic system is roughly the same in the longitudinal and transverse directions is indicating that the anisotropy produced by the magnetic field is not apparent in a system with very large hadronic masses. As we pointed out before, the field has to be larger than $10^{20}$ G to establish a quantum regime where the Landau quantization of the transverse momentum for protons becomes effective.

 \begin{figure}
\begin{center}
\begin{tabular}{ccc}
  \includegraphics[width=7cm]{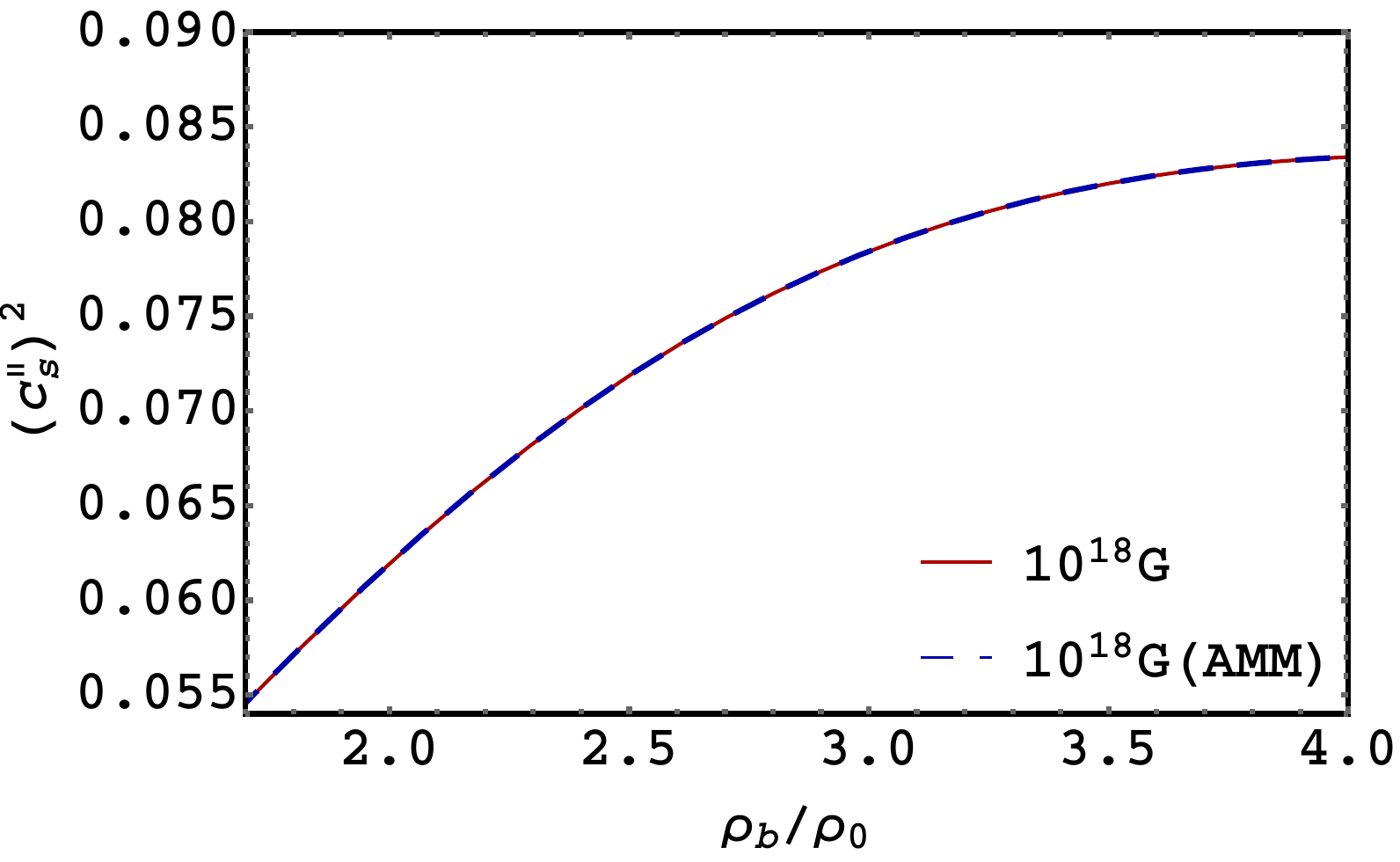} & \includegraphics[width=7cm]{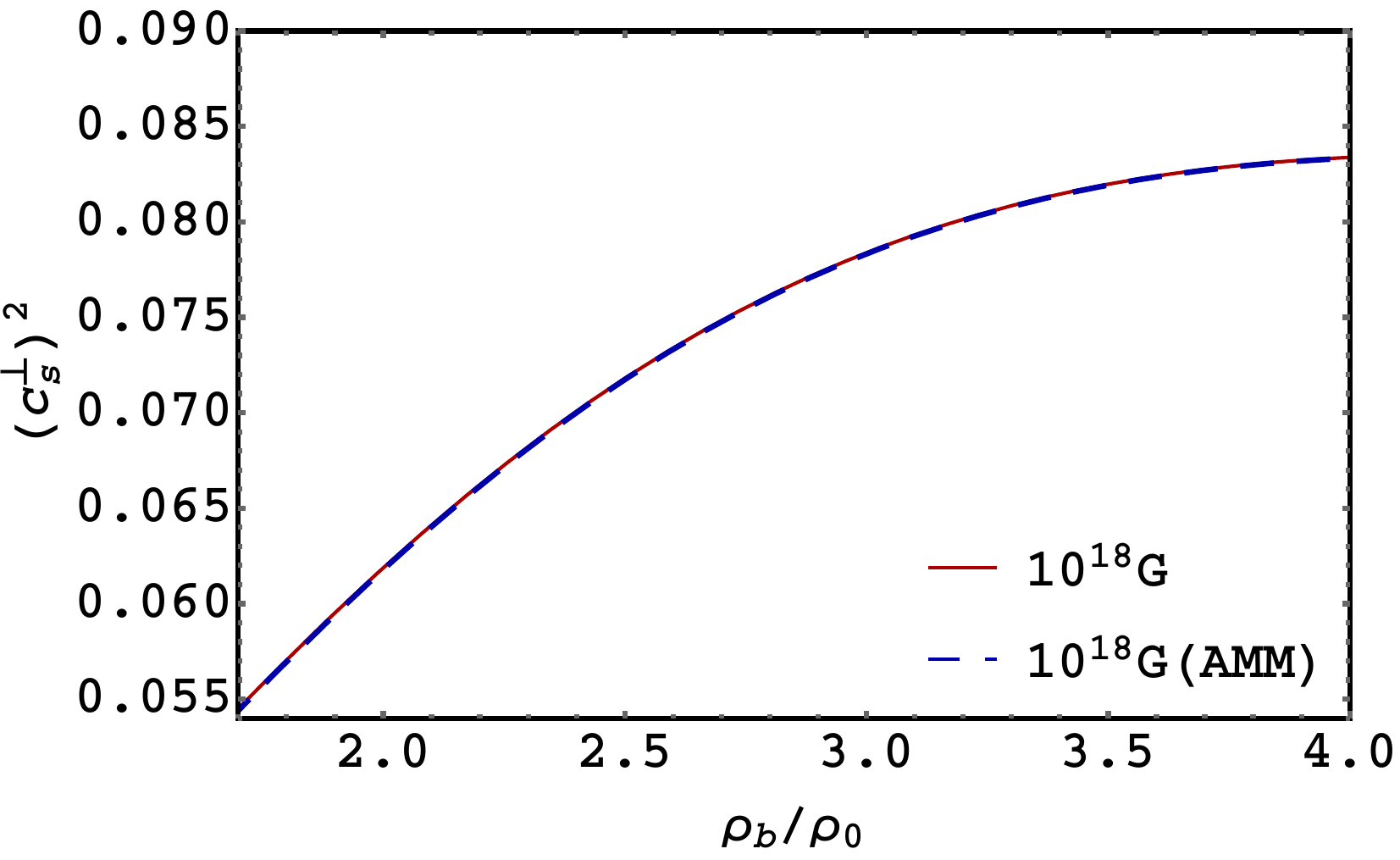}\\
(a) & (b)  \\
  \\
  \end{tabular}
    \end{center}
    \caption{(Color online) Parallel (a) and Perpendicular (b) SOS$^2$ in the hadron system versus normalized baryonic density $\rho_b/\rho_0$ ($\rho_0$ is the nuclear saturation density) at $10^{18}$G with the contribution of the magnetic field interaction with the neutron anomalous magnetic moment  (blue-dashed)
    and without that contribution (red-dotted).}
     \label{SOS-Hadrons-AMM}
\end{figure}

In Fig. \ref{SOS-Hadrons-AMM} the SOS of the magnetized hadronic phase is calculated including the effect of the magnetic field interaction with the neutron anomalous magnetic moment (B-N-AMM). From there, it can be observed that for hadrons the effect of the magnetic field is insignificant even up to field values of $\sim 10^{18}$ G independently of whether or not the interaction of the magnetic field with the N-AMM is taken into account. We should comment that when the B-N-AMM interaction is included in the many-particle thermodynamic potential, it was found \cite{neutrons, Lattimer-2000} that for magnetic fields that exceed the Fermi energy  ($\gtrsim 10^{18}$ G), the magnetic field contribution to the many-particle EOS becomes significant. That is, it was found that at those fields a complete spin polarization of the neutrons occurs, resulting in an overall stiffening of the many-particle EOS that overwhelms the softening induced by Landau quantization of the proton and electron spectra \cite{neutrons, Lattimer-2000}. Nevertheless, since the B-N-AMM interaction produces an increase of both the pressures and the energy density the two effects are compensated in such a way that no increase is reflected in the corresponding SOS's, as seen from Fig. \ref{SOS-Hadrons-AMM}.


Finally, it can be seen from Figs. \ref{SOS-Hadrons} and \ref{SOS-Hadrons-AMM}  that the parallel, as well as the transverse, SOS in the hadronic phase are well below the conformal limit.

 \subsection{ The SOS at intermediate densities: The MDCDW Phase}\label{section2-2}
 
 We consider now a region inside the inner core formed by quarks at intermediate densities $\sim 3 \rho_0$ and in the presence of a strong uniform magnetic field $\sim 10^{18}$ G, which is taken along the third spatial direction. For reasons already mentioned, it is reliable to consider that under these conditions the quarks are in the so-called MDCDW phase. In this phase, the quarks form an inhomogeneous chiral condensate characterized by two parameters: the condensate magnitude $m$ and modulation $q=b/2$ and having the following scalar and pseudoscalar  components
 \begin{equation}
\langle{\bar\psi}\psi\rangle=m\cos q_\mu x^\mu, \ \ \ \ \langle{\bar\psi}i\tau^3\gamma^5\psi\rangle=m\sin q_\mu x^\mu,
\end{equation}
The modulation is taking along the field direction, $q^\mu=(0,0,0,q)$ , since this configuration minimizes the system energy \cite{KlimenkoPRD82}. The MDCDW phase has different physical characteristics from the DCDW one. The flavor symmetry $SU(2)_L\times SU(2)_R$ of the DCDW phase is reduced to the subgroup $U(1)_L\times U(1)_R$ due to the coupling of quarks  of different electric charges with the magnetic field, and more importantly, the presence of the magnetic field induces a non-trivial topology in the MDCDW phase that gives rise to axion electrodynamics and to magnetoelectric effects \cite{Topological-Transport-1, Topological-Transport-2}. 

 In order to find the system SOS and keeping in mind that our purpose is to make our results applicable to NS, we only need to consider in the thermodynamic potential  the $T=0$ terms which depend on the chemical potential \cite{KlimenkoPRD82, Topological-Transport-1, Topological-Transport-2}
 \begin{equation} \label{thermo-pot}
\Omega_{q}=\Omega_{\rm anom}+\Omega_{\mu}
\end{equation}
with the anomalous contribution given by
\begin{equation}\label{Omega_anom}
\Omega_{\rm anom}=-N_c\frac{b}{2\pi^2}\sum_{f=u,d}|e_fB|\mu_f,
\end{equation} 
and the many-particle contribution equal to
\begin{equation} \label{omegamu}
\Omega_\mu=\Omega_{LLL}+\Omega_{hLL},
\end{equation}
with the LLL contribution given by
\begin{equation} \label{omegaLLL}
\Omega_{LLL}=-N_c\sum_{f=u,d}\frac{|e_fB|}{(2\pi)^2} [Q(\mu_f)+ m^2\ln (m/R(\mu_f) ]\theta [\mu_f-q/2-m],
\end{equation}

\begin{equation} \label{Qmu}
Q(\mu_f)=|q/2-\mu_f| \sqrt{(q/2-\mu_f)^2-m^2}, \quad R(\mu_f)=|q/2-\mu_f|+\sqrt{(q/2-\mu_f)^2-m^2},
\end{equation}
and the high Landau level (hLL) contribution by
\begin{equation} \label{omegamu-h}
\Omega_{hLL}=-N_c\sum_{f=u,d}\frac{b\mu_f}{2\pi^2}\int_{-\infty}^{\infty }dk \sum_{\xi, l>0}[\mu_f-E_l]\theta (\mu_f-E_l)|_{\epsilon=+}
\end{equation}
with energy spectra given by
\begin{eqnarray} \label{MDCDW-E}
   E_{0,\epsilon}&=&\epsilon \sqrt{m^2+p^2}+b, \quad \epsilon=\pm, \l=0 \nonumber
\\
E^f_{l,\xi,\epsilon}&=&\epsilon \left [ \left (\xi\sqrt{m^2+p^2}+b \right )^2+2|e_f B| l \right ]^{1/2}, \quad \epsilon=\pm, \xi=\pm, l=1,2,3,...
\end{eqnarray}

The anomalous contribution $\Omega_{\rm anom}$ comes from the regularization of the LLL part ensuring that the thermodynamic potential is independent of $b$ when $m=0$ \cite{KlimenkoPRD82} and it is a consequence of the asymmetry with respect to zero energy of the energy spectrum in the LLL (\ref{MDCDW-E}). Moreover, the LLL thermodynamic potential (\ref{omegaLLL}) corresponds to the intermediate density region where $\mu > m+q/2$ \cite{Topological-Transport-1, Topological-Transport-2}, which is the one we want to consider.

The order parameters $m$ and $b$ should be determined from the gap equations 
\begin{equation}\label{Gap-Eq}
\frac{\partial\Omega_q}{\partial m}=\frac{\partial\Omega_q}{\partial b}=0,
\end{equation}
what indicates their dynamical origin.

We assume that the system is beta equilibrated 
\begin{equation}\label{quarkchempotentials}
	\mu_u=\mu-\frac{2}{3}\mu_e, \quad \mu_d=\mu+\frac{1}{3}\mu_e.
\end{equation}
and neutral 
\begin{equation}\label{Q-neutrality-1}
	\frac{\partial (\Omega_q+\Omega_e)}{\partial \mu_e}=0.
\end{equation}
$\Omega_e$ is the electron thermodynamic potential, 
\begin{equation}\label{Omega-e-2}
\Omega_e=-\frac{\lvert{eB}\rvert}{\left(2\pi\right)^2}\displaystyle\sum_{l=0}^{l_{max}}g_l\left\{\mu_e\sqrt{\mu_e^2-M_l^2}-M_l^2\ln\left\lvert\frac{\mu_e+\sqrt{\mu_e^2-M_l^2}}{M_l}\right\rvert\right\},
\end{equation}
where $\mu_e$ is the electric chemical potential, $g_l$ is the spin degeneracy with $g_0=2-\delta_0^l$, $M_l=\sqrt{\mu_e^2+2\lvert{eB}l\rvert}$, and

	\begin{equation}
		l_{max}=\left\lfloor\frac{\mu_e^2-m_e^2}{2\lvert{eB}\rvert}\right\rfloor.
	\end{equation}
 As is known, the electron contribution is needed in the two-flavor quark system to reach electric neutrality. 

 From the Heaviside functions in (\ref{omegamu-h}) we get the constraint for the Landau number
 \begin{equation}\label{Constraint}
l_{max}\leqslant \mu_f^2 / 2|e_fB|.
\end{equation}
Notice from (\ref{Q-neutrality-1}) that $\mu_e$ is a dependent parameter of $\mu$ and $B$.
We will consider quark chemical potentials with values around $\mu=320$ MeV, which corresponds to intermediate densities $\sim 3\rho_0$. Now, to constrain the quarks into the LLL we need a magnetic field with a value at least one order larger than $\mu_f^2/2|e_f|$. For the value of $\mu$ under consideration, it implies a magnetic field strength with a magnitude beyond $10^{18}$ G, that is, beyond the maximum value we want to consider. But, considering acceptable values for the inner magnetic field $\sim 10^{18}$ G, in the range of $\mu$ values under consideration the constraint (\ref{Constraint}) implies that only a few Landau levels will contribute depending on the quark flavor. Hence, as follows, we make some calculations that will then be used in the numerical calculation of the SOS.

 \begin{figure}
\begin{center}
\begin{tabular}{ccc}
  \includegraphics[width=8cm]{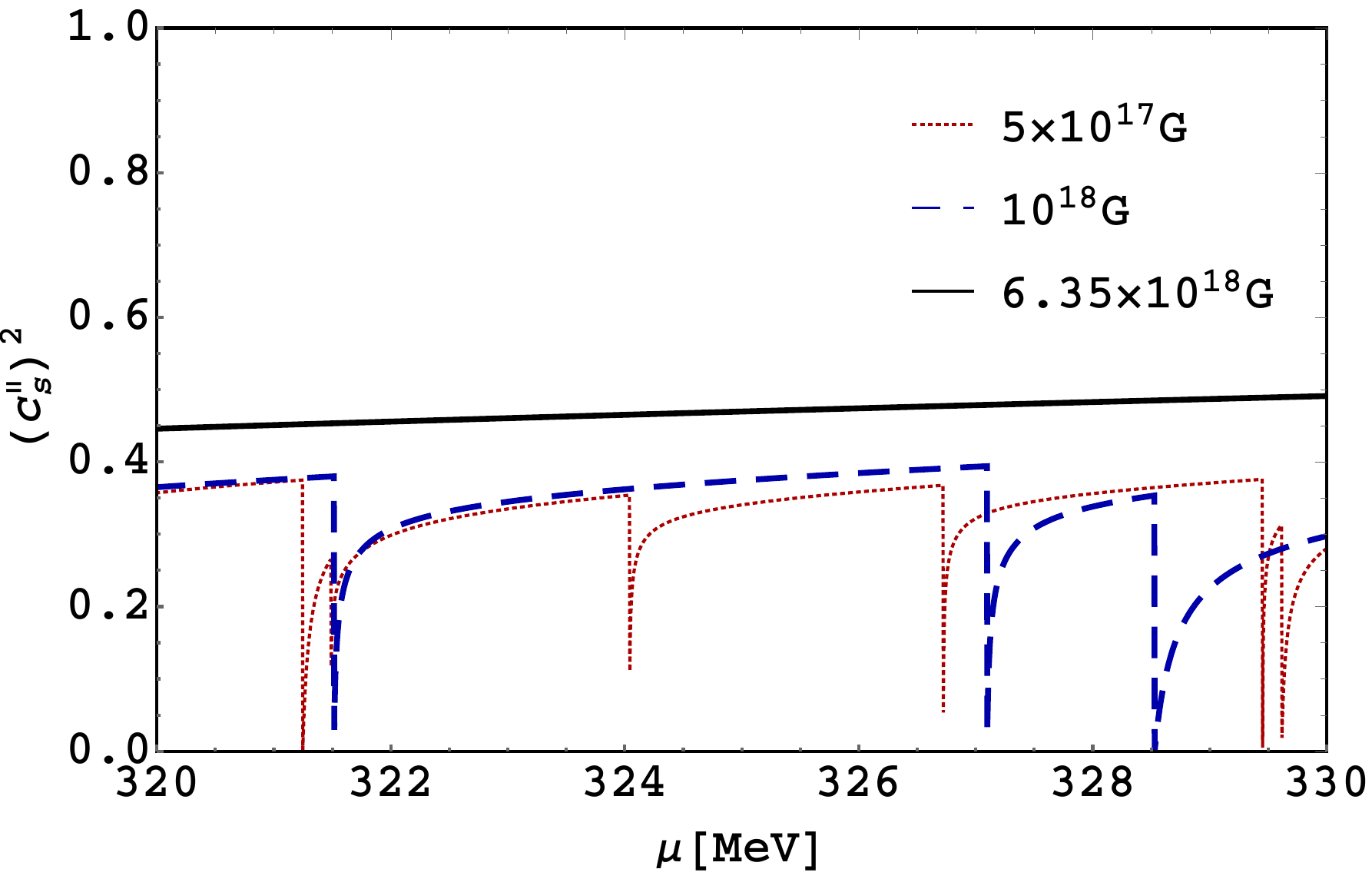} & \includegraphics[width=8cm]{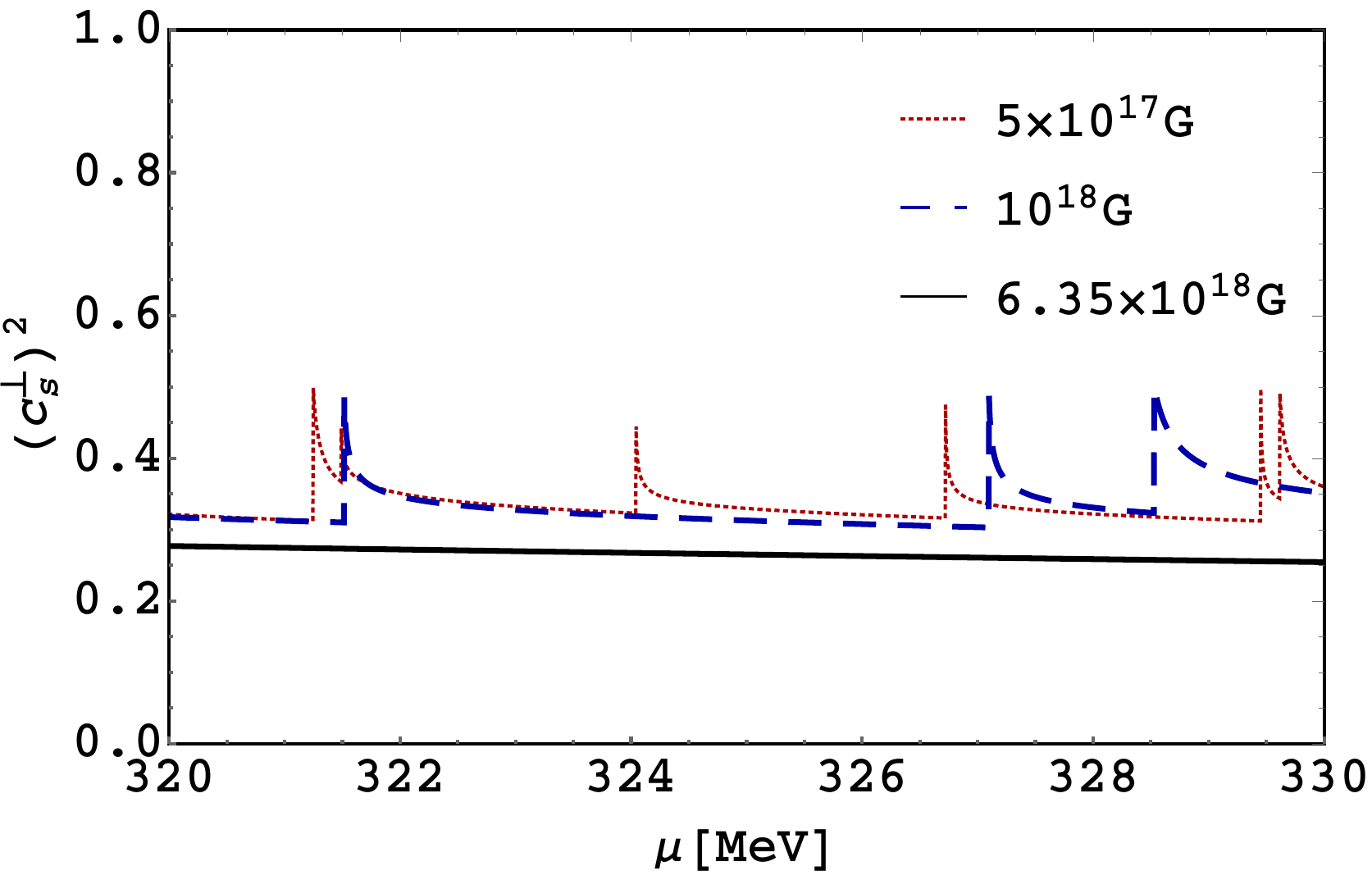}\\
(a) & (b)  \\
  \\
  \end{tabular}
    \end{center}
    \caption{(Color online) Parallel (a) and Perpendicular (b) SOS$^2$ in the MDCDW phase versus baryonic chemical potential corresponding to intermediate densities
    for different applied magnetic fields.}
     \label{MDCDW-Par-Per}
\end{figure}

\subsection{MDCDW thermodynamic potential in the $m^2 <\mu^2, B$ region}

Taking into account that in the considered region, $m^2 <\mu^2, B$, we can neglect the mass in (\ref{thermo-pot}), as can be seen in Ref. \cite{Will}, we find
 \begin{equation}\label{Omega-l0}
\Omega_{l=0} = \Omega_{anom}+\Omega_{LLL}\simeq -  \frac{N_c} {(2\pi)^2}   \sum_{f=u,d}|e_fB|[q\mu_f+\mu_f^2 -q\mu_f]=- \frac{N_c} {(2\pi)^2}  \sum_{f=u,d}|e_fB|\mu_f^2
\end{equation}

For  higher LL's the quark contribution becomes
\begin{equation}\label{PPerr-l1}
\Omega_{l>0} = -2 N_c \sum_{f=u,d}\frac{|e_fB|}{(2\pi)^2} \sum_{\xi=\pm} \int_0^\infty dk_3 \left [ \mu_f - \sqrt{(k_3+\xi q/2)^2 + 2 |e_fB|l} \right ] \theta[\sqrt{\mu_f^2-2|e_fB|l}-\xi q/2-k_3]
\end{equation}

After integration it is found
\begin{eqnarray}\label{Omega-l1}
\Omega_{q} &=& \Omega_{anom}+\Omega_{LLL}+\Omega_{l>0} \simeq -\frac{N_c\lvert{eB}\rvert}{\left(2\pi\right)^2} \sum_{f=u,d}\Bigg\{ \mu_f^2+2 \sum_{l=1}^{l_f^{max}}\Bigg[\mu_f\sqrt{\mu_f^2-2|e_fBl|} \nonumber
\\
&-&2\lvert{eB}\rvert\ln \left\lvert\frac{\mu_f-\sqrt{\mu_f^2-2|e_fBl|}}{ \sqrt{2|e_fBl|}}\right\rvert \Bigg]\Bigg\},
\end{eqnarray}

where

\begin{equation}
l_f^{max}=\left\lfloor\frac{\mu_f^2}{2\lvert{eB}\rvert}\right\rfloor.
\end{equation}\label{L-Max}

In Fig. \ref{MDCDW-Par-Per} we plot the parallel and perpendicular components of SOS$^2$ in the MDCDW phase for different magnetic fields in the density region of interest. Notice the jumps corresponding to the de Hass-van Alphen oscillations, which take place when a new LL is filled. That is, each jump in the graphs means that as we increase the chemical potential, keeping the magnetic field constant, new LL's enter to contribute. See that if the field decreases, more LL's contribute at a given density. 

Another interesting effect that takes place in this phase is that while the parallel SOS decreases with $\mu$, the perpendicular one increases. It is produced by the fact that the perpendicular pressure depends on the magnetization (\ref{Pressures-EOS-1}), which in this case has an anomalous component that increases with $\mu$, as can be seen from (\ref{Omega_anom}). Finally, comparing the results of Fig. \ref{MDCDW-Par-Per} with those of Fig. \ref{SOS-Hadrons}  we see that the SOS peaks at intermediate densities as we expected (i.e. the parallel and perpendicular components of SOS at intermediate densities are one order higher than those in the hadronic phase).

\begin{figure}[t]
\begin{center}
\includegraphics[width=0.5\textwidth]{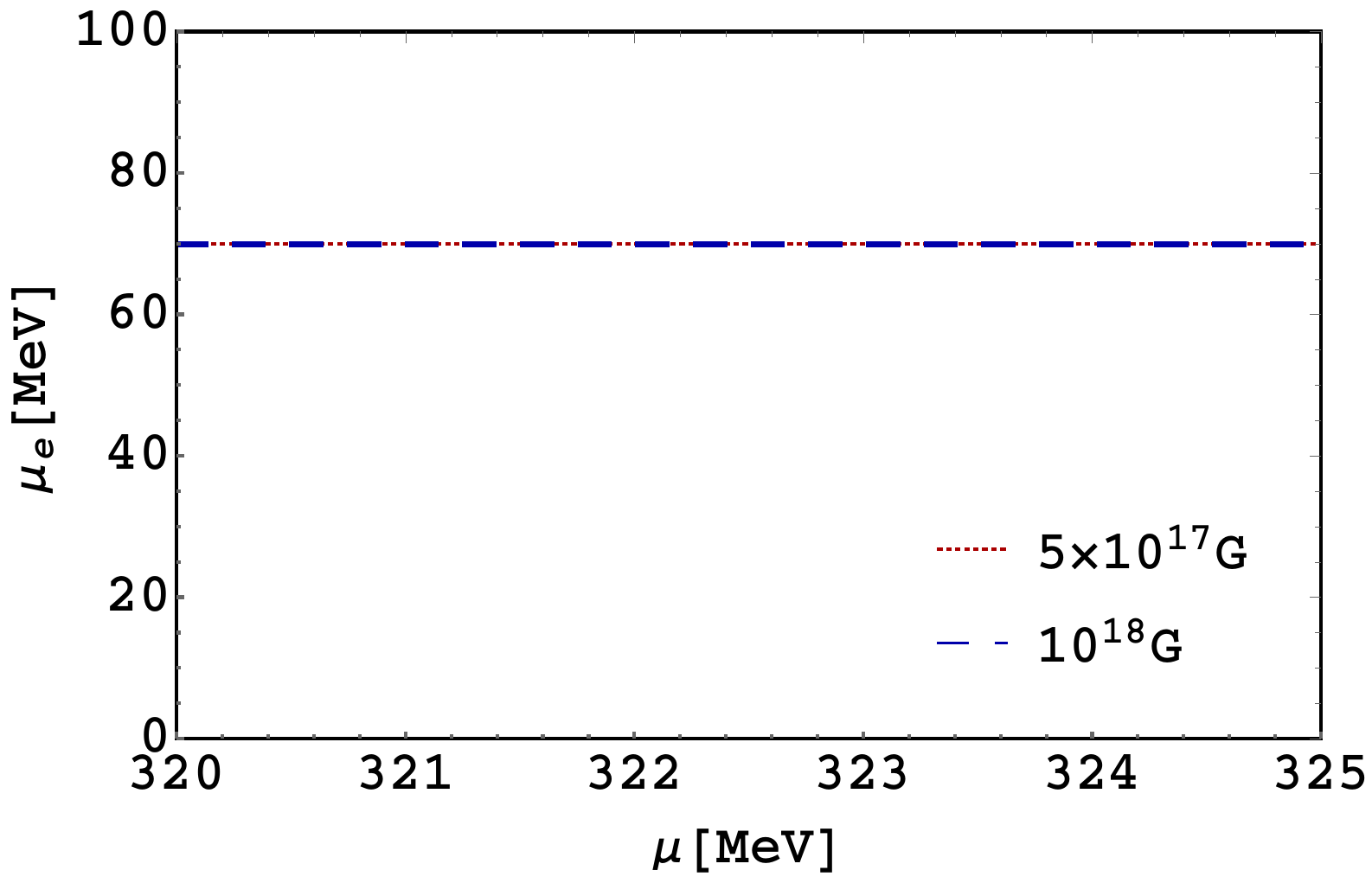}
\caption{(Color online)  Electric chemical potential $\mu_e$ versus baryon chemical potential $\mu$ in the MDCDW phase
    for different applied magnetic fields.}
\label{muVmu_e}
\end{center}
\end{figure}

We call attention that although the electric chemical potential is one order smaller than the baryonic chemical potential, as seen from Fig. \ref{muVmu_e}, it can affect the number of LL's, which can participate at some given value of the magnetic field as seen from the constraint (65).

\subsection{ The SOS at high densities: The Magnetized Free-Quark Phase}\label{section2-1} 

Due to asymptotic freedom, at sufficiently high densities  the quarks behave as free particles.  To describe the magnetized free-quark phase we use the MIT bag model \cite{Bag} in the presence of a uniform magnetic field directed  along the third spatial direction. In this model,  the quarks are taken to be free inside of an effective "bag". The effects of the bag are formally realized in the model by adding (subtracting) a fixed constant to the energy density (pressure). We consider a system of up and down quarks, and electrons (as already mentioned, the electrons are needed in a two-flavor system to ensure electric neutrality) to be  immersed in a uniform background magnetic field $\vec{B}=B\hat{z}$.

Since we are considering that the baryonic chemical potential is the leading parameter in this region, in finding the system thermodynamic potential we can use the WFA where the Landau-level sum is replaced by the integrals in the transverse momenta. In the WFA, the finite-density one-loop thermodynamic potential for each quark flavor is arrived at by taking $m_e{\to}m_q$, $e\to{q_q}$, $\mu_e\to{\mu_q}$ in $\Omega_e$ in Eq. (\ref{Omega-e}), and multiplying by an overall factor of $3$ to account for the color degeneracy. Hence, the fermion thermodynamic potential in this phase  is
\begin{equation}\label{ThermoPotentialQuarks-1}
	\Omega^{Q}_f=\Omega_u+\Omega_d+\Omega_e,
\end{equation}
where
\begin{equation}\label{ThermoPotentialQuarks-2}
	\begin{aligned}
	  &\Omega_u=\frac{-1}{8\pi^2}\left\{\left(2{\mu_u}^4-5{m_u}^2{\mu_u}^2\right)\sqrt{1-\left(\frac{m_u}{\mu_u}\right)^2}+\left(3{m_u}^4+\frac{8}{9}\left(eB\right)^2\right)\ln\left[\frac{{\mu_u}+\sqrt{{\mu_u}^2-{m_u}^2}}{{m_u}}\right]\right\},
\\
&\Omega_d=\frac{-1}{8\pi^2}\left\{\left(2{\mu_d}^4-5{m_d}^2{\mu_d}^2\right)\sqrt{1-\left(\frac{m_d}{\mu_d}\right)^2}+\left(3{m_d}^4+\frac{2}{9}\left(eB\right)^2\right)\ln\left[\frac{{\mu_d}+\sqrt{{\mu_d}^2-{m_d}^2}}{{m_d}}\right]\right\}.
\end{aligned}
\end{equation}
are respectively the $u$ and $d$ one-loop thermodynamic potentials in the WFA, and $\Omega_e$ is the electron thermodynamic potential given in (\ref{Omega-e}).  We use for the current quark masses $m_u=m_d=5.5$MeV and $\mu_q$ denotes the chemical potential of flavor $q$. 

Then, using (\ref{energy-density-1})-(\ref{Pressures-EOS-1}) and after adding the bag constant we get the EOS of the magnetized system given by
\begin{equation}\label{PressuresQS}
	\begin{aligned}
	\varepsilon&={\Omega}_f^{Q}+\displaystyle\sum_{i=u,d,e}\mu_i{\rho}_i^{Q}+\frac{B^2}{2}+Bag,
\\
P^{Q}_{\perp}&=-{\Omega}_f^Q-BM_f^Q+\frac{B^2}{2}-Bag,
\\
P^{Q}_{\parallel}&=-{\Omega}_f^{Q}-\frac{B^2}{2}-Bag.
	\end{aligned}
\end{equation}
with $M_f^Q=-\partial{\Omega^Q}/\partial{B}$ being the quark system magnetization, and  the quark number densities are given by
\begin{equation}\label{numberdensitiesQS}
\begin{aligned}
	{\rho}^Q_u&=\frac{1}{\pi^2}\left[\left({\mu_u}^2-m_u^2\right)^{\frac{3}{2}}+\frac{(eB)^2}{9\sqrt{{\mu_u}^2-m_u^2}}\right],
\\
	{\rho}^Q_d&=\frac{1}{\pi^2}\left[\left({\mu_d}^2-m_d^2\right)^{\frac{3}{2}}+\frac{(eB)^2}{36\sqrt{{\mu_d}^2-m_d^2}}\right].
\end{aligned}
\end{equation}
The electron number density is the same as in (\ref{numberdensitiesHS}). 

We assume that the system is beta equilibrated, so satisfying Eq. (\ref{Q-neutrality}),
and neutral 
\begin{equation}\label{Q-neutrality}
	\frac{\partial \Omega^Q_f}{\partial \mu_e}=\frac{2}{3}\rho_u-\frac{1}{3}\rho_d-\rho_e=0
\end{equation}

The baryonic charge density is given by 
\begin{equation}
	\rho^{Q}_b=-\frac{\partial \Omega^Q_f}{\partial \mu}=\frac{1}{3}\rho_u+\frac{1}{3}\rho_d.
\end{equation}

\begin{figure}
\begin{center}
\begin{tabular}{ccc}
\includegraphics[width=8cm]{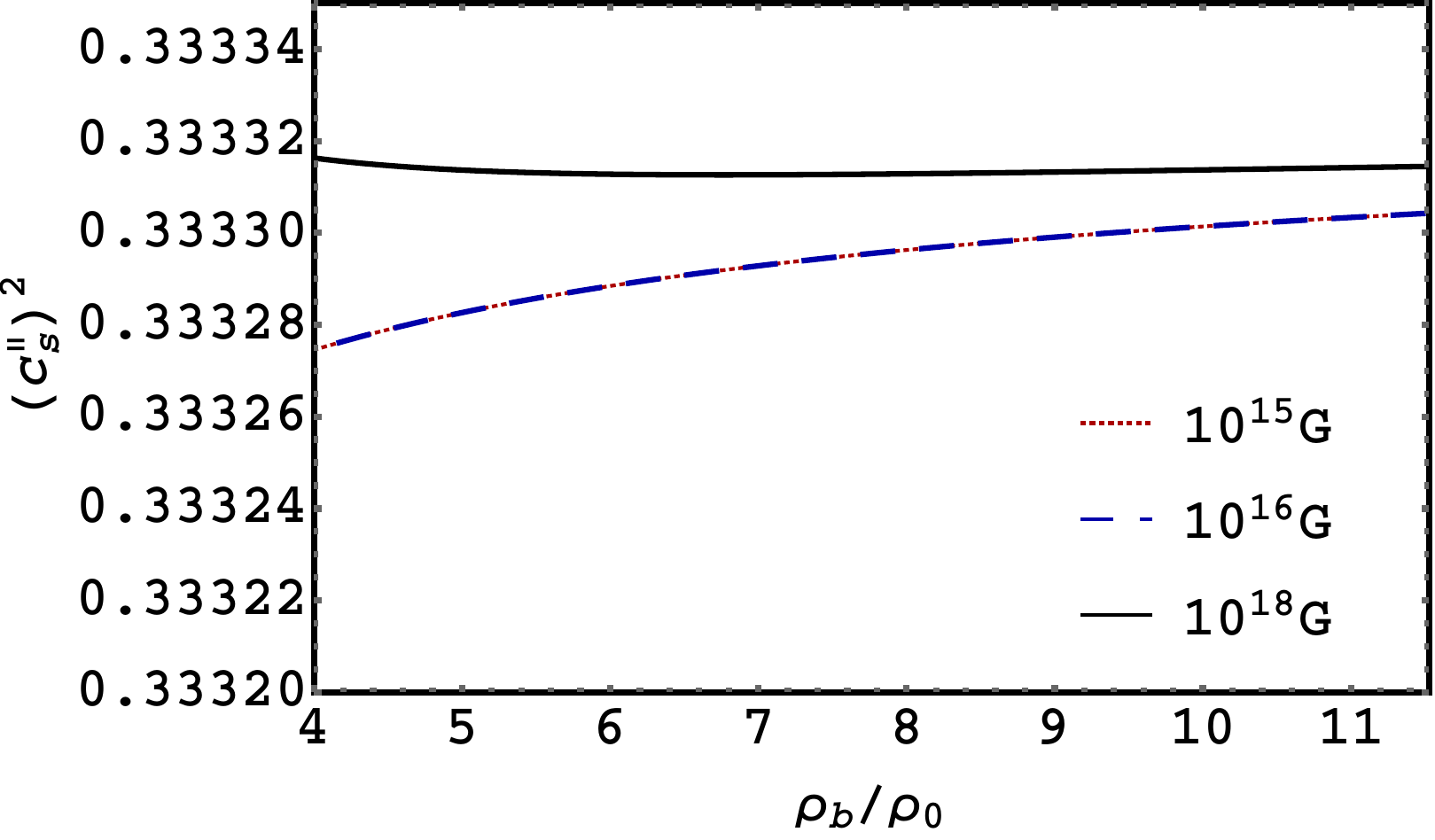} & \includegraphics[width=8cm]{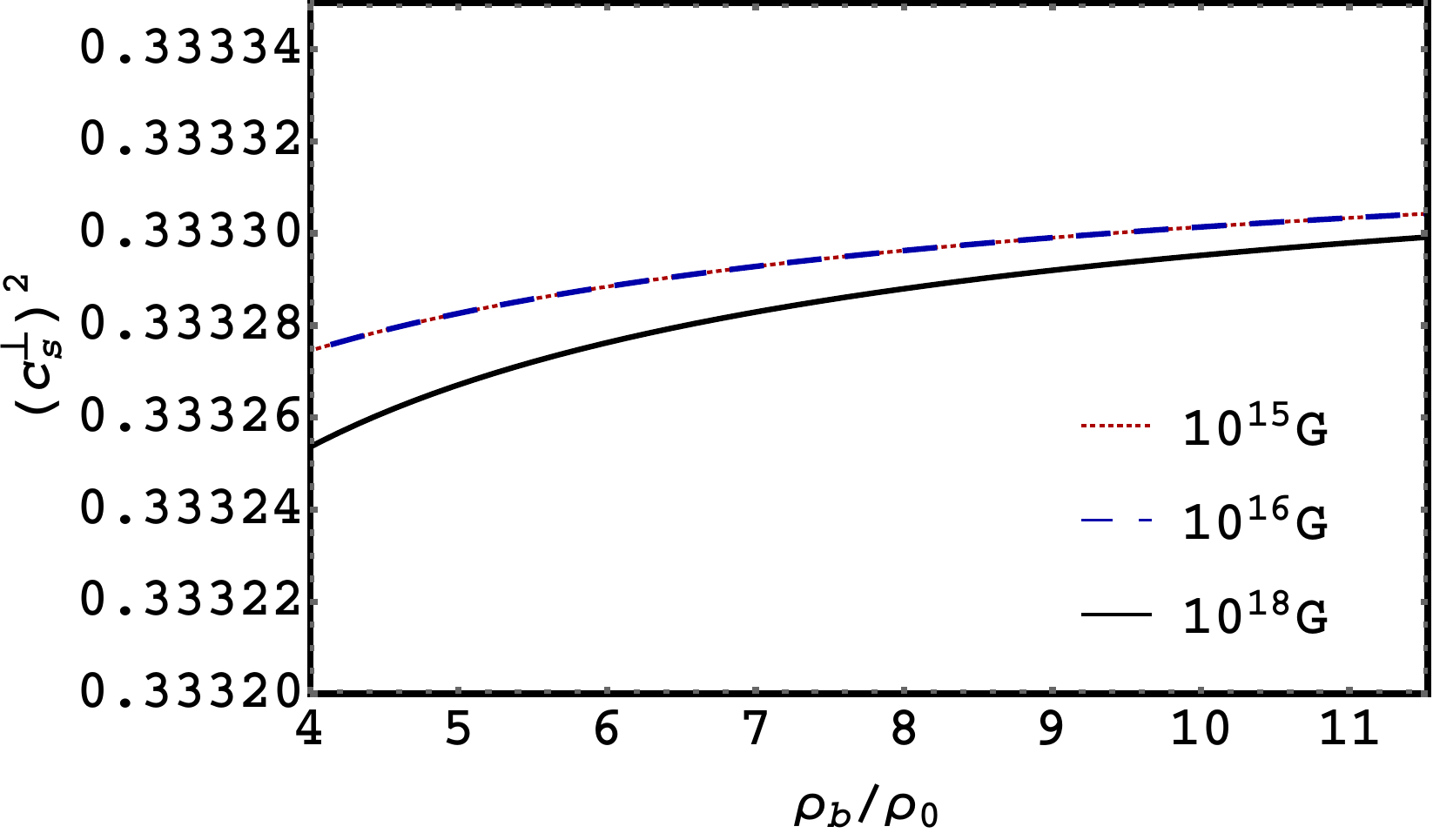} \\
  \\
  \end{tabular}
    \end{center}
    \caption{(Color online) Parallel (\textbf{Left}) and Perpendicular (\textbf{Right}) SOS$^2$ quark systems versus normalized baryonic charge density $\rho_b/\rho_0$ at magnetic field strength values of $10^{15}$G (red-dotted), $10^{16}$G (blue-dashed), and $10^{18}$G (black-solid).}
     \label{Quarks-SOSPlot}
\end{figure}

Using Eqs. (\ref{velocidades}) and (\ref{SOS}), we calculate the SOS versus density for this case and the results appear in Fig. \ref{Quarks-SOSPlot}.  As seen from Fig. \ref{Quarks-SOSPlot} the parallel and transverse SOS at those densities approach the conformal limit from below as the density increases, corroborating what had been predicted  \cite{Bedaque}.

 \section{Concluding remarks}\label{Conclusions}
 
 In Ref. \cite{Nature2020} it was found by using a model-independent calculation together with astrophysical observation data, that NS with masses corresponding to 1.4 $M_{\odot}$  are compatible with nuclear model calculations, but those with $M\approx 2M_{\odot}$ have characteristics indicative of the existence of deconfined quark matter in their cores. On the other hand, as discussed in \cite{Bedaque}, the existence of such maximum masses $\sim  2M_{\odot}$, implies that the SOS as a function of the energy density gets a peculiar shape. It should rise from small values, $c_s^2 < 1/3$, to reach a maximum surpassing the conformal limit, $c_s^2\geqslant 1/3$, and then decreasing to a minimum value to finally reach the conformal limit, $c_s^2=1/3$, from below at higher densities. This scenario will respond to the energy density profile inside a massive, $\sim  2M_{\odot}$, hybrid star, whose internal structure corresponds to different matter phases  starting from a hadronic phase in the outer core region with a low value of $c_s$, passing through an intermedium density region with deconfined quark matter in the external part of the inner core (that can be presumably formed by an inhomogeneous quark-hole phase as the MDCDW one we are considering), where the SOS can surpass the conformal limit, to finally reach a conformal free quark phase at sufficiently high densities at the smallest radii. 
 
 In order to make this scenario plausible, we have found that a strong stellar magnetic field can play a significant role. Specifically, the relevance that the LLL contribution can have in the different phases will determine  the values $c_s$ can get across the radial stellar profile. The relevance of the LLL is determined by the fact that the fermions in this ground state are subjected to a spatial dimensional reduction (d=1), which as we explained in the Introduction, and then proved in Sections III and IV-B, is a key element in producing $c_s$ values beyond the conformal limit. 
 
The change in the $c_s$ profile inside the NS  is due to the interrelation between three parameters:  the particle masses, the magnetic field and the baryonic chemical potential.
  In this regard, due to the large hadron masses, for magnetic field strengths acceptable for NS interior (i.e. $\lesssim 10^{18}$ G) and moderate densities ($m^2 \gg B, \mu^2$) the possibility to confine a significant number of particles into  the LLL is impossible. This implies that at low energy densities $c_s < 1/3$, as we have shown in Section IV-A. In the other extreme in density, when the baryonic chemical potential becomes the leading parameter, $\mu^2 \gg B, m^2$, all the Landau levels equally contribute, minimizing the effect of the LLL, and producing a pressure $p \sim \mu^4$, which, as we already noticed in Fig. \ref{Quarks-SOSPlot}  gives rise to an SOS, which approaches from below the conformal limit $c_s^2 = 1/3$. In the intermediate density region occupying the zone inside the core closer to the hadron-quark phase transition boundary and having a moderate  baryonic density ($B> m^2, B\gtrsim \mu^2$), the role of the LLL can be predominant at fields $\sim 10^{18}$ G, since at those fields and densities only a few Landau levels will be filled, which makes the role of the LLL more significant. In this phase, there are two other elements that also help in rising the value of the SOS: first, that at those intermediate densities the particle masses given by the magnitude of the inhomogeneous chiral condensate are negligibly small as compared with $\sqrt{B}$ \cite{KlimenkoPRD82, Will}, and second, that the anomaly contribution to the thermodynamic potential characteristic of this phase (\ref{Omega_anom}), makes a positive contribution to the  pressure linearly proportional to the magnetic field and baryonic chemical potential, which contributes to stiffer the EOS and hence to increase $c_s$. 
   
Thus, we conclude that our investigation has shown that the stellar magnetic field can play an important role in producing the $c_s$ profile expected to be characteristic of very massive NS  \cite{Bedaque}. Moreover, in our approach the realization at intermediate densities of a quark phase as the MDCDW phase was instrumental in producing a peak in the SOS. Hence, this result is calling attention on the  close bound between stars with high masses  $\sim  2M_{\odot}$  and the existence of quark matter in the core, as had been already indicated in \cite{Nature2020}.

Nevertheless, to complete the physical scenario by  checking if the expected SOS profile \cite{Bedaque} in the presence of a magnetic field is realized after building a hybrid star composed of nuclear and quark matter,
it is first necessary to find how in the range of baryonic densities that characterize the outer and inner regions of the star core, the three phases we have considered can take place and if such a star structure will be able to reach through the TOV equations the maximum mass $\sim  2M_{\odot}$. We should mention that a preliminary study in that direction was already done in  \cite{InhStars} showing that the corresponding EOS supports stars with masses around $2M_{\odot}$ for values of the magnetic field that are in accordance with those inferred from magnetar data. But there, only the hadronic and MDCDW phases were considered and for the last one a simpler version of the model was worked out in which different quark flavors can be completely decoupled by
neglecting the instanton term in the interaction Lagrangian. Thus, to completely  accomplish this task a consideration of a neutral MDCDW phase, which includes the instanton term \cite{Instanton} that accounts for the $U (1)_A$
anomaly of QCD is necessary. This is a project in progress. Moreover, a similar study to that done in \cite{PhaseTrans-B} for the anisotropic first-order phase transitions between the phases under consideration in the presence of a magnetic field will be also required to find the corresponding critical values of the baryonic chemical potentials.

 Another important outcome of our investigation is the reported anisotropy in the speed of sound in the presence of a magnetic field. In the magnetized medium, the speed of sound is split with different values along and transverse to the magnetic field direction. This is a consequence of the splitting of the EOS in a magnetized medium \cite{magnetizedfermions, PhaseTrans-B}. We have shown that in the strong-field limit, when $B \gtrsim \mu^2, m^2$, the transverse speed velocity approaches zero, indicating that when the quarks are confined into the LLL there is no medium available to transmit a pressure perturbation in that direction; while in this limit the speed of sound  along the field direction reaches the causal limit $c_s^\| \approx 1$, which is an indication of the large increase of the fluid stiffness along the field lines.
 
Since the SOS is a measure of the stiffness of the EOS, the fact that the MDCD phase has a larger SOS than the magnetized hadronic phase, indicates that a star with a magnetized quark phase can reach a larger mass. This result is in agreement with a recent finding \cite{Nature2020} showing that the heaviest neutron stars, with masses $\sim 2 M_\odot$, should have deconfined quark-matter in their cores and also with the mentioned results of \cite{InhStars}, where stellar masses $\sim 2M_{\odot}$ can be reached by hybrid star having quark matter in the MDCD phase at the core.
It is likely that future observations will provide more compelling evidence in favor of models with EOS where $c_s$ goes above the conformal limit \cite{Comp-Arguments}.

\acknowledgments
This work was supported in part by NSF grant PHY-2013222.

\newpage 

\end{document}